\documentclass[universe,article,accept,moreauthors]{Definitions/mdpi}

\usepackage[utf8]{inputenc}  
\usepackage{amsmath}
\usepackage{amsfonts}
\usepackage{siunitx}
\usepackage{longtable}
\usepackage[normalem]{ulem}
\usepackage{comment}
\usepackage{booktabs} 
\usepackage{siunitx}  
\usepackage{threeparttable} 

\firstpage{1} 
\pubvolume{1}
\issuenum{1}
\articlenumber{0}
\pubyear{2026}
\copyrightyear{2026}
\externaleditor{Firstname Lastname} 
\datereceived{12 January 2026} 
\daterevised{5 March 2026} 
\dateaccepted{17 March 2026} 
\datepublished{ } 

\Title{Connecting Meteorite Spectra to Lunar Surface Composition Using Hyperspectral Imaging and Machine Learning}

\Author{Fatemeh Fazel Hesar $^{1,2}$\orcidB{},
Mojtaba Raouf  $^{2,3,4,}$*\orcidA{},
Amirmohammad Chegeni $^{5,6}$\orcidC{},
Peyman Soltani $^7$, \linebreak  
Bernard Foing $^{2,4,8}$,
Elias Chatzitheodoridis $^9$,
Michiel J. A. 
de Dood $^7$ and Fons J. Verbeek $^1$
}

\AuthorNames{Fatemeh Fazel Hesar, Mojtaba Raouf, Amirmohammad Chegeni, Peyman Soltani, Bernard Foing, Elias Chatzitheodoridis, Michiel J. A. de Dood, Fons J. Verbeek}

\address{$^{1}$ \quad Leiden Institute of Advanced Computer Science, Leiden University, P.O. Box 9504,  
 2300 RA 
 Leiden, The Netherlands; f.fazel.hesar@liacs.leidenuniv.nl (F.F.H.); f.j.verbeek@liacs.leidenuniv.nl (F.J.V.) 
 \\
$^{2}$ \quad ILEWG LUNEX-EuroSpaceHub EuroMoonMars Earth-Space Innovation Wassenaar, \linebreak  Leiden \& Noordwijk,
 The Netherlands; foing@strw.leidenuniv.nl\\
$^{3}$ \quad Department of Space Engineering, Delft University of Technology, 2629 HS Delft, The Netherlands\\
$^{4}$ \quad Leiden Observatory, Leiden University, {P.O. Box 9513,} 2300 RA Leiden, The Netherlands \\

$^{5}$ \quad Dipartimento di Fisica e Astronomia ``G. Galilei'', Università di
Padova, Via Marzolo 8, 35131 Padova, Italy; amirmohammad.chegeni@unipd.it or amirmohammad.chegeni@pd.infn.it 
 \\
$^{6}$ \quad INFN-Padova, Via Marzolo 8, 35131 Padova, Italy\\

$^{7}$ \quad  Huygens-Kamerlingh Onnes Laboratory, Leiden University, Niels Bohrweg 2, \linebreak  2333 CA Leiden, The Netherlands; soltani@physics.leidenuniv.nl (P.S.); \linebreak  dood@physics.leidenuniv.nl (M.J.A.d.D.)
 \\
$^{8}$ \quad ERA Chair of Space Photonics , NSP Fotonika, Latvia University, Riga,\\ 
$^{9}$ \quad Department of Geological Sciences, School of Mining and Metallurgical Engineering, National Technical University of Athens, Iroon Polytechniou 9, Zografou Campus, 15773 Athens,
 Greece; eliasch@metal.ntua.gr\\

}

\corres{Correspondence: mojtaba.raouf@gmail.com or raouf@strw.leidenuniv.nl
}

\abstract{We present an innovative, cost-effective framework integrating laboratory Hyperspectral Imaging (HSI) of the Bechar 010 Lunar meteorite with ground-based lunar HSI and supervised Machine Learning (ML) to generate high-fidelity mineralogical maps. A \SI{3}{\milli\metre} thin section of Bechar 010 was imaged under a microscope with a \SI{30}{\milli\metre} focal length lens at \SI{150}{\milli\metre} working distance, using 6x binning to increase the signal-to-noise ratio, producing a data cube (X $\times$ Y $\times$ $\lambda$ = $791 \times 1024 \times 224$, \SI{0.24}{\milli\metre} $\times$ \SI{0.2}{\milli\metre} resolution) across \SIrange{400}{1000}{\nano\metre} (224 bands, \SI{2.7}{\nano\metre} spectral sampling, \SI{5.5}{\nano\metre} full width at half maximum spectral resolution) using a Specim FX10 camera.
Ground-based lunar HSI was captured with a Celestron 8SE telescope (\SI{3}{\kilo\metre}/pixel), yielded a data cube ($371 \times 1024 \times 224$). Solar calibration was performed using a Spectralon reference ({99}\% reflectance {<2}\% error) ensured accurate reflectance spectra. 
A Support Vector Machine (SVM) with a radial basis function kernel, trained on expert-labeled spectra, achieved {93.7}\% classification accuracy (5-fold cross-validation) for olivine ({92}\% precision, {90}\% recall) and pyroxene ({88}\% precision, {86}{\%} recall) in Bechar 010. LIME analysis identified key wavelengths (e.g., \SI{485}{\nano\metre}, {22.4}\% for M3; \SI{715}{\nano\metre}, {20.6}\% for M6) across 10 pre-selected regions (M1 to M10), indicating olivine-rich (Highland-like) and pyroxene-rich (Mare-like) compositions. SAM analysis revealed angles from \SI{0.26}{\radian} to \SI{0.66}{\radian}, linking M3 and M9 to Highlands and M6 and M10 to Mares. K-means clustering of Lunar data identified 10 mineralogical clusters ({88}\% accuracy), validated against Chandrayaan-1 Moon mineralogy Mapper ($\rm M^3$) data (\SI{140}{\metre}/pixel, \SI{10}{\nano\metre} spectral resolution). A novel push-broom HSI approach with a telescope achieves 0.8 arcsec resolution for lunar spectroscopy, inspiring full-sky multi-object spectral mapping.}


\begin{document}
\section{Introduction}
\label{sec:introduction}

Lunar surface mineralogy provides critical insights into geological processes such as basaltic heterogeneity and mantle-derived magmatism, which have shaped the Moon’s mare and highland regions \cite{SURKOV2020, Bhatt2019}. Basaltic heterogeneity in mare regions, driven by diverse volcanic processes, results in varied mineralogical compositions, notably olivine and pyroxene, as observed in regions like Aristarchus Plateau and Mare Imbrium \cite{Head1976}. The Lunar surface is 
 extremely complex, shaped by an uncountable number of impacts by natural projectiles of diverse compositions, resulting in extensive mixing at different scales~\cite{Young2016, Warren2005}. These breccias exemplify the Moon's continuous exposition to such bombardment \cite{WarrenTaylor2014}. Lunar meteorites provide ground-truth samples of the Lunar surface. Bechar 010, classified as a Lunar feldspathic breccia \cite{Gattacceca2024}, contains anorthositic clasts with minor olivine, pyroxene, and recrystallized glass, making it a valuable analog for studying both highland and mare-derived lithologies within a single sample. Recent advances extend ML to Lunar meteorite mineral classification and mechanical property prediction \cite{PenaAsensio2024}, complementing spectral-focused approaches.

Hyperspectral Imaging (HSI) has revolutionised Lunar mineralogy mapping by enabling detailed spectral characterization across contiguous wavelength bands, revealing the Moon's geological history through volcanic processes, crustal differentiation, and impact-driven resurfacing \cite{Head1976, McSween2006, Pieters2009}. Olivine and pyroxene, key mafic minerals in Lunar basalts and highlands, are critical for their petrogenetic significance and distinct spectral signatures in the 400--1000~nm range \cite{Cloutis1986}. Recent advances in Machine Learning (ML), such as Support Vector Machines (SVMs) and unsupervised clustering, have enhanced the precision of mineralogical identification in hyperspectral data, supporting NASA's Artemis program by enabling high-resolution mineralogical mapping of the Lunar crust \cite{Sabat2020}.

Lunar meteorites, such as Bechar 010, provide accessible analogs to Lunar terrains, with analyses revealing olivine-rich compositions that mirror mare lavas' volcanic \linebreak  origins~\cite{Head1976, Corley2018, Elkins2003}. These findings align with remote sensing studies, including a study that used Chandrayaan-1 Moon mineralogy Mapper ($\rm M^3$) data to map ilmenite and TiO$_2$ abundances, emphasizing the role of spectral data in Lunar mineralogy~\cite{SURKOV2020}. Similarly mapped elemental abundances (Fe, Ca, Mg) using $\rm M^3$ data minimized space weathering artifacts to provide global context for localized meteorite studies, supporting our focus on olivine and pyroxene identification \cite{Bhatt2019}.

In our previous paper, \cite{Fazel2025}, we explored the mineralogical composition of volcanic samples akin to lunar materials, emphasising olivine and pyroxene. Utilizing HSI from 400 to 1000 nm, we analyzed data cubes of samples from the volcanic island Vulcano, Italy, categorizing them into nine regions of interest. The study employed various unsupervised clustering algorithms, revealing that hierarchical clustering provided the most reliable results, with K-means also demonstrating strong performance in identifying olivine predominance linked to Lunar-like volcanic processes. 
Prior studies provide a foundation for our work,  using absorption features in $\rm M^3$ data to identify ilmenite, olivine, and pyroxene, paralleling our spectral analysis approach \cite{SURKOV2020}. A study  demonstrated the effectiveness of K-means clustering for lunar mineralogical classification \cite{Thoresen2024}, complementing our use of SVM-based methods, while another study \cite{Zhang2021} refined techniques to distinguish olivine and pyroxene based on diagnostic spectral features \cite{Plaza2011}. This study aims to integrate laboratory HSI of the Bechar 010 meteorite with ground-based and orbital data, using ML to produce high-resolution mineralogical maps that enhance our understanding of Lunar surface composition and geological evolution. 

Lunar meteorites, such as Bechar 010, a feldspathic breccia, offer critical insights into the mineralogical and compositional diversity of the Lunar surface, serving as proxies for studying lunar highland and mare regions without direct in situ sampling~\cite{korotev2005}. This study uses hyperspectral analysis in the visible to near-infrared range 
(\SIrange{400}{1000}{\nano\meter}) to characterize Bechar 010’s spectra, employing the Spectral Angle Mapper (SAM) algorithm to quantify similarities with Lunar spectra and Local Interpretable Model-agnostic Explanations (LIMEs) to identify key wavelengths driving Support Vector Machine (SVM) classification~\cite{pieters1993, ribeiro2016}. 
By associating the dominant spectral wavelengths with key mineral phases, including olivine, pyroxene, and anorthite, our analysis highlights the heterogeneous composition of Bechar 010 and its affinity with Lunar lithologies. These findings not only shed light on the complex mineralogical assemblages within the meteorite but also provide insights into Lunar geological processes and the provenance of meteorite material~\cite{Burns1993}.

Despite the success of orbital HSI, as demonstrated in Chandrayaan-1's M$^3$ mapping Fe$^{2+}$ spectral absorption features near 1 and 2~\textmu m wavelengths to identify olivine and pyroxene \cite{Mustard2011, Pieters2009}, ground-based HSI remains underutilized due to atmospheric interference and calibration challenges \cite{Zhang2021}. Lunar meteorites like Bechar 010 offer a unique opportunity to bridge laboratory and remote sensing data, providing accessible analogs to Lunar terrains \cite{Elkins2003}. This study introduces a novel framework that integrates laboratory HSI of Bechar 010 with ground-based lunar HSI and SVM-based classification to generate mineralogical maps comparable to orbital datasets, advancing Lunar geological analysis.

This paper is structured as follows: Section~\ref{sec:methodology} describes the the data and methodology; \textls[-25]{Section~\ref{sec:results} presents the findings; and Sections~\ref{sec:discussion_conclusion} and \ref{sec:conc} discuss the implications and~conclusions.}

\section{Data and Methodology}
\label{sec:methodology}

This study employs a comprehensive framework to map Lunar mineralogy, integrating laboratory and ground-based HSI with advanced ML and interpretability techniques. The methodology uses the Bechar 010 Lunar meteorite and Lunar surface observations, validated against Chandrayaan-1 Moon mineralogy Mapper (M$^3$) reflectance data, to quantify olivine and pyroxene distributions across diverse terrains such as the Aristarchus Plateau, Mare, and Highland basins. Spectral data were acquired across a 400--1000~nm range using dual HSI systems, ensuring consistency for cross-domain analysis following our Paper-I for Volcanic samples \cite{Fazel2025}. 
For ground-based observation, atmospheric turbulence and seeing conditions were considered during data acquisition, as they can affect spatial resolution. 
To mitigate these effects, observations were conducted under optimal seeing conditions, with an estimated seeing-limited resolution of approximately 1.6 arcsecond.

\begin{figure}[H]
\includegraphics[width=0.9\linewidth]{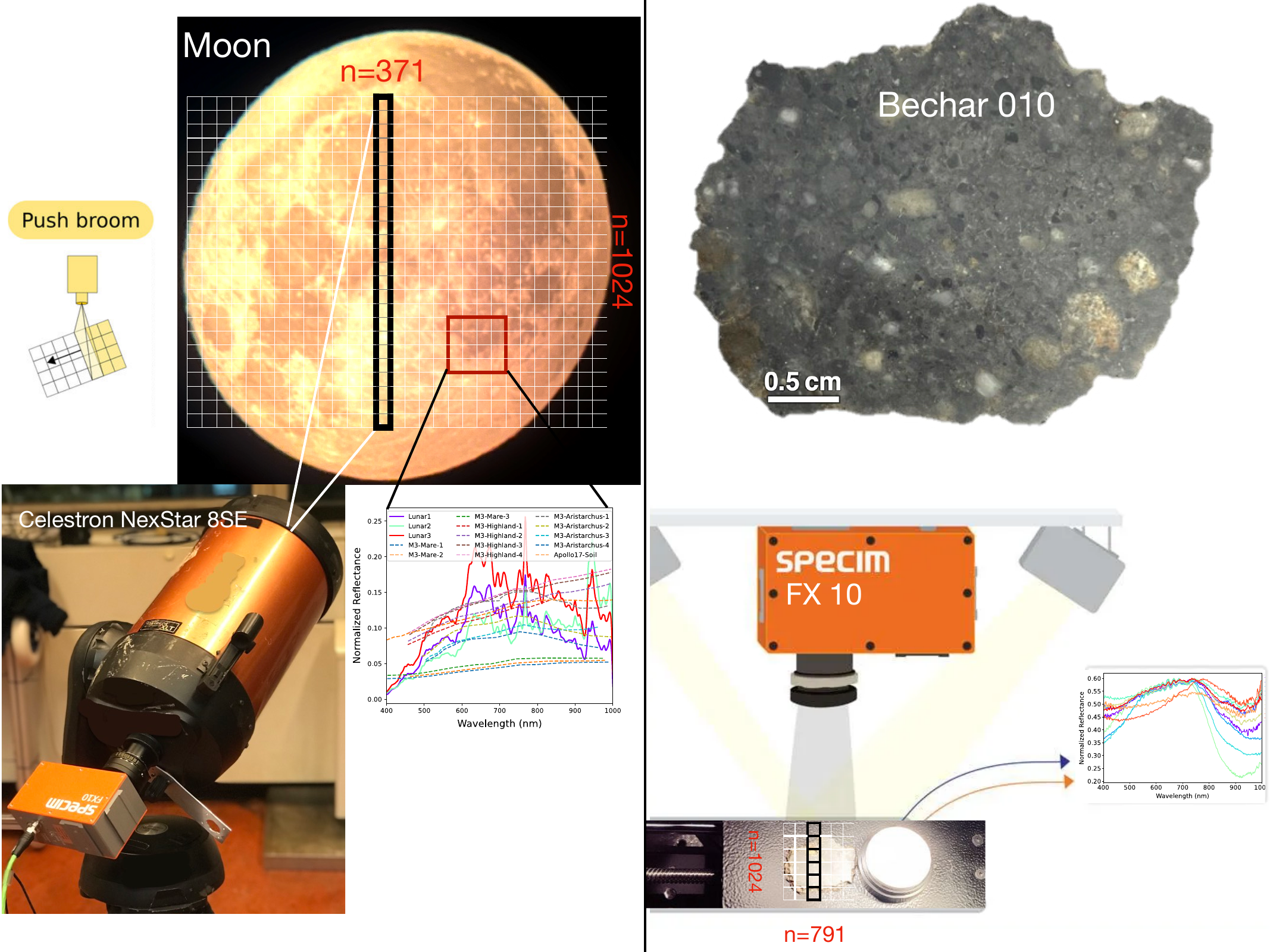}
\caption{(\textbf{Right}): Experimental setup for HSI of Bechar 010 meteorite, featuring the Specim FX10 camera, halogen lamps, and motorized scanner covered 791 pixels. An example of the spectral data for the regions of interest (ROIs) is also provided, highlighted with colored lines. (\textbf{Left}): Celestron 8SE Nexstar telescope to create a hyperspectral image of the Moon via a scan using a custom push-broom spectrometer (400--1000~nm) technique to create the spectral image map. The HSI leveraged the Moon’s relative motion concerning Earth to perform the scan. The scan direction covered 371 pixels, which was sufficient to map the Lunar surface based on the Moon’s angular velocity (approximately 0.004°/s). An example of the spectral data for selected region is highlighted with the colored lines.}
\label{fig:Setup_all}
\end{figure}

The HSI system employs a Specim FX10 camera with a CMOS sensor operating in a controlled dark room environment. It captures 1024 spatial pixels across 224 spectral bands (400--1000 nm, FWHM 5.5 nm), using halogen lamps for consistent illumination and a motorized linear stage for precise sample positioning. Preprocessing includes dark current correction and reflectance calibration with a Spectralon white reference (99\% reflectance) to ensure accurate reflectance data. The setup, illustrated in Figure \ref{fig:Setup_all}, enables the identification of subtle spectral features associated with minerals like olivine and pyroxene. As shown in the right panel of Figure \ref{fig:Setup_all}, the Bechar 010 meteorite, a feldspathic breccia from Algeria (2022), is a \SI{3}{\milli\metre} thick, \SI{3}{\centi\metre} $\times$ \SI{3.5}{\centi\metre} slab from a \SI{681}{\gram} fragment \cite{Gattacceca2024}. It contains anorthositic clasts, olivine, pyroxene, and recrystallized glass, resembling Lunar highlands and impact ejecta \cite{Elkins2003}. A Specim FX10 hyperspectral camera, paired with a microscope imaging system with a 30 mm focal length lens at a working distance of approximately \SI{150}{\milli\metre}, captured reflectance spectra in a controlled laboratory environment to generate a high-spectral and -spatial-resolution hyperspectral map. The sample was scanned in XY under the microscope lens, resulting in a three-dimensional data cube ($791 \times 1024 \times 224$) with each pixel containing the reflectance intensity at a specific wavelength. The spatial resolution is approximately \SI{0.24}{\milli\metre} in the x-direction and \SI{0.2}{\milli\metre} in the y-direction, determined by the macroscope’s field of view and the camera’s pixel array.

\textls[-25]{HSI of the Moon was conducted using a Celestron 8SE telescope (203 mm aperture, 2032 mm focal length) at an urban site in Leiden in the Netherlands (52.16° N, 4.49° E, 6 m altitude), near the Leiden Observatory, established in 1633 
(\url{https://www.universiteitleiden.nl/en/science/astronomy}).
 Frames were captured over three nights under clear skies, with temperatures of \mbox{5--10 °C} and wind speeds of 5--15 mph.} The ASI1600MM camera used high gain settings, consistent 5 ms exposures across spectral bands, and sensor cooling slightly above the \mbox{5--10~°C} ambient minimum to minimize noise. Preprocessing included solar calibration using a Spectralon reference (99\% reflectance) at a solar zenith angle of 51°31$'$09.7$"$, measured with a 1.648 ms exposure, followed by dark subtraction and reflectance normalization via \( \text{Reflectance} = (\text{Raw} - \text{Dark}) / (\text{White} - \text{Dark}) \). Outlier removal via statistical thresholding eliminated anomalous pixels, and a Savitzky--Golay filter with a window length of 31 and polynomial order of 5 was applied to smooth spectral profiles while preserving absorption features critical for mineralogical analysis.
\subsection{Classification Techniques for Hyperspectral Data Analysis}
We present a novel approach to Lunar surface mineralogy by integrating HSI and ML to classify and map olivine and pyroxene distributions. A laboratory-based dataset derived from the feldspathic breccia Lunar meteorite Bechar 010 serves as ground truth for supervised Support Vector Machine (SVM) classification. This classifier is then applied to remotely acquired HSI data from the Lunar surface using a custom-built telescope equipped with a pushbroom hyperspectrometer. 
We applied SVM classification to hyperspectral data from ten Regions of Interests (ROIs), integrating a weighted probability approach with binary classification based on class probabilities. To interpret the model, we used LIME as an explainable AI tool, which identified the five most influential wavelengths guiding SVM decision-making. These diagnostic features were then linked to mineralogical compositions, enabling the mapping of spectral signatures to lunar-relevant phases. This approach emphasizes the value of combining machine learning with physical interpretability, demonstrating how rigorous evaluation methods can strengthen algorithm validation and enhance the reliability of mineralogical classification.

\subsubsection*{ SVM Model Configuration and Hyperparameter Optimization}
The Radial Basis Function (RBF) kernel was selected for the SVM classifier because hyperspectral mineral classification involves non-linearly separable classes in high-dimensional spectral space. The RBF kernel maps data into an infinite-dimensional feature space via the Gaussian function $K(\mathbf{x}_i, \mathbf{x}_j) = \exp(-\gamma \|\mathbf{x}_i - \mathbf{x}_j\|^2)$, enabling effective separation of spectrally similar minerals such as olivine and pyroxene without explicit feature engineering. Unlike linear or polynomial kernels, RBF handles complex, overlapping absorption features characteristic of mineral mixtures.

The two key hyperparameters---regularization parameter $C$ and kernel width\linebreak   $\gamma$---were optimized via a grid search over logarithmic scales ($C \in \{0.1, 1, 10, 100, 1000\}$; $\gamma \in \{1, 0.1, 0.01, 0.001, 0.0001\}$), evaluated using 5-fold cross-validation accuracy. The optimal combination ($C = 100$, $\gamma = 0.01$) achieved the highest mean cross-validation accuracy of {93.7}\%. A high $C$ value enforces strict margin separation (risking overfitting on noisy spectra), while a low $\gamma$ ensures that the kernel captures broad spectral features rather than noise. The pipeline proceeded as follows: (1) spectral preprocessing including normalization and continuum removal; (2) stratified 80/20 train--test splitting preserving class proportions; (3) grid search with 5-fold cross-validation on training data; (4) final evaluation on the held-out test set.

The SVM model was trained on a total of approximately 5600 labeled spectra extracted from the 10 ROIs (M1--M10) within the Bechar 010 data cube ($791 \times 1024 \times 224$). Each ROI contains between 400 and 800 pixel spectra, selected based on expert visual identification of mineral phases under the microscope, cross-referenced with known reflectance characteristics. Training spectra were labeled into seven olivine--pyroxene mixture categories (O100\%, O90\%-P10\%, \ldots, P100\%) using reference spectra from the ROMA database \cite{Mandon2022}. A learning curve analysis showed that classification accuracy saturated at approximately 3000 spectra (reaching {93}\% accuracy), with marginal improvement beyond 4000 spectra, confirming that the training set size is sufficient for stable SVM performance. The 5-fold cross-validation further confirmed model stability, with standard deviations of $\pm${1.2}\% in accuracy across folds.

\subsection{Lunar Clustering Method and Compared to Moon Mineralogy Mapper (M$^3$)}
We compare K-means clustering maps of Lunar surface data. 
The resulting mineralogical maps exhibit high consistency with Moon mineralogy Mapper ($\rm M^3$) data, particularly across selected regions of the Aristarchus Plateau and Mare deposits. Comparative analysis with previous studies \cite{Bhatt2019, Mustard2011} further supports the robustness and geological significance of our results. This approach establishes a pipeline linking terrestrial analogs, meteoritic data, and Lunar remote sensing, significantly contributing to planetary spectroscopy and future Lunar exploration strategies.

The Spectral Angle Mapper (SAM) algorithm was used to compare hyperspectral data from Bechar 010 and Lunar surface observations (\SIrange{400}{900}{\nano\metre}) with reference spectra from lunar regolith analogs~\cite{Mandon2022}. SAM was selected because it computes spectral similarity based on the angle between spectral vectors, making it insensitive to differences in illumination intensity and albedo---a critical advantage when comparing laboratory spectra (Bechar 010, under halogen illumination) with remotely sensed Lunar spectra (under solar illumination). Unlike Euclidean distance-based measures, SAM focuses on spectral shape rather than absolute reflectance, which is particularly important for cross-domain comparisons where calibration differences may affect reflectance magnitudes 
\cite{Kruse1993}. The spectral angle, \(\alpha\), is calculated as:

\begin{equation}
\alpha = \cos^{-1}\left( \frac{\mathbf{t} \cdot \mathbf{r}}{\|\mathbf{t}\| \|\mathbf{r}\|} \right).
\end{equation}
where \(\mathbf{t}\) is a vector that contains the target spectrum from the sample (e.g., Bechar 010 or Lunar regions) and the vector \(\mathbf{r}\) contains the reference spectrum from known Lunar analogs (e.g., olivine, pyroxene). This angle quantifies spectral similarity, with smaller values indicating closer matches \cite{Kruse1993}. A threshold of \SI{0.3}{\radian}, consistent with prior hyperspectral classification studies, was adopted to ensure robust matches. Lunar Mare spectra showed matches with basaltic references (\(\alpha = \SI{0.61}{\radian}\)), with discrepancies due to basaltic heterogeneity. Bechar 010 aligned with Lunar analogs, with \(\alpha = \SI{0.26}{\radian}\) for olivine-rich regions (M3, M9) and \(\alpha = \SI{0.61}{\radian}\) for pyroxene-rich regions (M6, M10).
To generate a high-resolution RGB image of lunar hyperspectral data, we developed an enhanced image processing pipeline using Python
 with the \textit{PIL}
 \cite{PIL}, \textit{numpy} \cite{NumPy}, \textit{scipy} \cite{SciPy}, and \textit{matplotlib}~\cite{matplotlib}~libraries. Hyperspectral data from a FITS file was loaded, and three spectral bands corresponding to red (650 nm), green (550 nm), and blue (450 nm) wavelengths were extracted using nearest-neighbor wavelength indexing. Each band underwent a multi-step enhancement process to improve smoothness and resolution. First, Gaussian smoothing ({$\sigma=1.5$} pixel unit) was applied to reduce noise. The bands were then normalized to the range [0, 255] and enhanced for brightness (offset = 15) and contrast (factor = 1.3). To enhance edge details, a sharpening convolution was applied twice using a 3 $\times$ 3 kernel (\textit{[[$-$1, $-$1, $-$1], [$-$1, 9, $-$1], [$-$1, $-$1, $-$1]]}). For super-resolution, each band was upscaled by a factor of 4 using cubic spline interpolation, followed by light Gaussian smoothing ({$\sigma=0.5$} pixel unit) to minimize interpolation artifacts.

\subsection{
Highlighting Influential Wavelengths in ML Decision-Making via LIME}

ML models have demonstrated great potential in analysing spectral data, particularly for tasks such as classification and regression, especially in the field of \linebreak  astrophysics~\cite{Bhatt2019,mlspectral2,mlspectral3,mlspectral4,Fazel2025b,Ghaziasgar2025}. However, their black-box nature often makes some data untrustworthy and leads to misinterpretations, especially when decisions rely on many individual wavelengths. Explainable Artificial Intelligence (XAI) methods \cite{xai1}, such as LIME (Local Interpretable Model-agnostic Explanations) \cite{ribeiro2016}, address this challenge by identifying which specific wavelengths most strongly influence the model's decisions. By applying LIME to spectral models, we can gain insight into the physical meaning behind the machine decisions, and make sure that the model is based on scientifically relevant features. This improves both the transparency and reliability of ML in spectroscopy-driven applications.

LIME is an explainable AI technique designed to make the decision-making process of ML models transparent. It operates on the principle of building a simpler, interpretable surrogate model that locally mimics the behaviour of the original model around a specific input. To do this, LIME perturbs the original input data and observes how the model's decisions change. This surrogate model shows which input features are most influential for the decision of that particular instance. By focusing on the local neighborhood around the input, LIME offers an intuitive way to understand individual predictions without needing access to the internal structure of the original model \cite{xai3}.
Each region in our dataset consists of a lot of wavelength-reflectance features (e.g., there are 6300 features for region 3), representing the spectral response across a wavelength range. To interpret the predictions made by the SVM classifier, we apply the LIME framework. For a given sample, LIME perturbs its high-dimensional feature vector (corresponding to the flattened spatial--spectral representation of the hyperspectral cube) by slightly modifying values at different wavelength positions. This generates synthetic spectra in the vicinity of the original sample. These perturbed samples are then evaluated using the trained SVM model, and the corresponding prediction probabilities are recorded. This procedure enables LIME to assess how the model’s output varies in response to changes at specific wavelength features. LIME perturbs its spectral feature vector by slightly modifying reflectance values at different wavelength positions, thereby generating synthetic spectra in the vicinity of the original sample. These perturbed spectra are then evaluated using the trained SVM model, and the corresponding prediction probabilities are recorded. This procedure allows LIME to assess how the model’s output changes in response to variations across specific wavelength features. Using the output of these perturbations, LIME fits an interpretable linear surrogate model that approximates the SVM’s local decision boundary near the input sample. The weights (coefficients) of this linear model indicate the relative importance of each wavelength band for the prediction. To make the explanation concise and interpretable, LIME ranks all spectral features and selects the top five bands with the highest absolute weights. These represent the most influential wavelengths driving the classification. While this subset does not capture the full variability of the spectrum, it highlights the dominant spectral regions guiding the model’s decision, offering a compact and human-readable explanation of classifier behavior.

\section{Results}
\label{sec:results}
\subsection{Mineralogical Classification of Bechar 010}
\label{sec:bechar_classification}
As shown in the right panel of Figure \ref{fig:Setup_all}, the hyperspectral cube is generated using a  push-broom line-scan technique with the following steps: (1) The meteorite slab is placed on a motorized stage under the camera, with the slit aligned along the y-direction. (2) The stage moves the sample in the y-direction at a constant rate, synchronized with a 3.9 times slower than the camera’s frame rate (The Specim
 FX10 camera has a \SI{10}{\milli\metre} $\times$ \SI{47}{\micro\metre} slit size. For a \SI{35}{\milli\metre} sample covering the slit with 3.5$\times$ magnification, the line image width is $3.5 \times \SI{47}{\micro\metre} = \SI{0.147}{\milli\metre}$, requiring $\SI{30}{\milli\metre} / \SI{0.147}{\milli\metre} \approx 202$ bands to cover a \SI{30}{\milli\metre} sample. Acquiring 791 bands indicates a scanning speed 3.9 times slower than optimal (791/202). Averaging from 791 to 202 bands and recording parameters (e.g., exposure time, gain) is considered for improved performance, capturing 791 pixels across the 30~mm width reflectance spectra for a 35 mm $\times$ 0.147 mm. (3) The stage completes 1024~steps to cover the \SI{35}{\milli\metre} height, producing a $791 \times 1024 \times 224$ data cube. (4) The data are processed to map mineralogical distributions, enabling detailed analysis of the meteorite’s composition and texture, which is comparable to Lunar analogs.

\begin{figure}[H]
\includegraphics[width=0.35\linewidth]{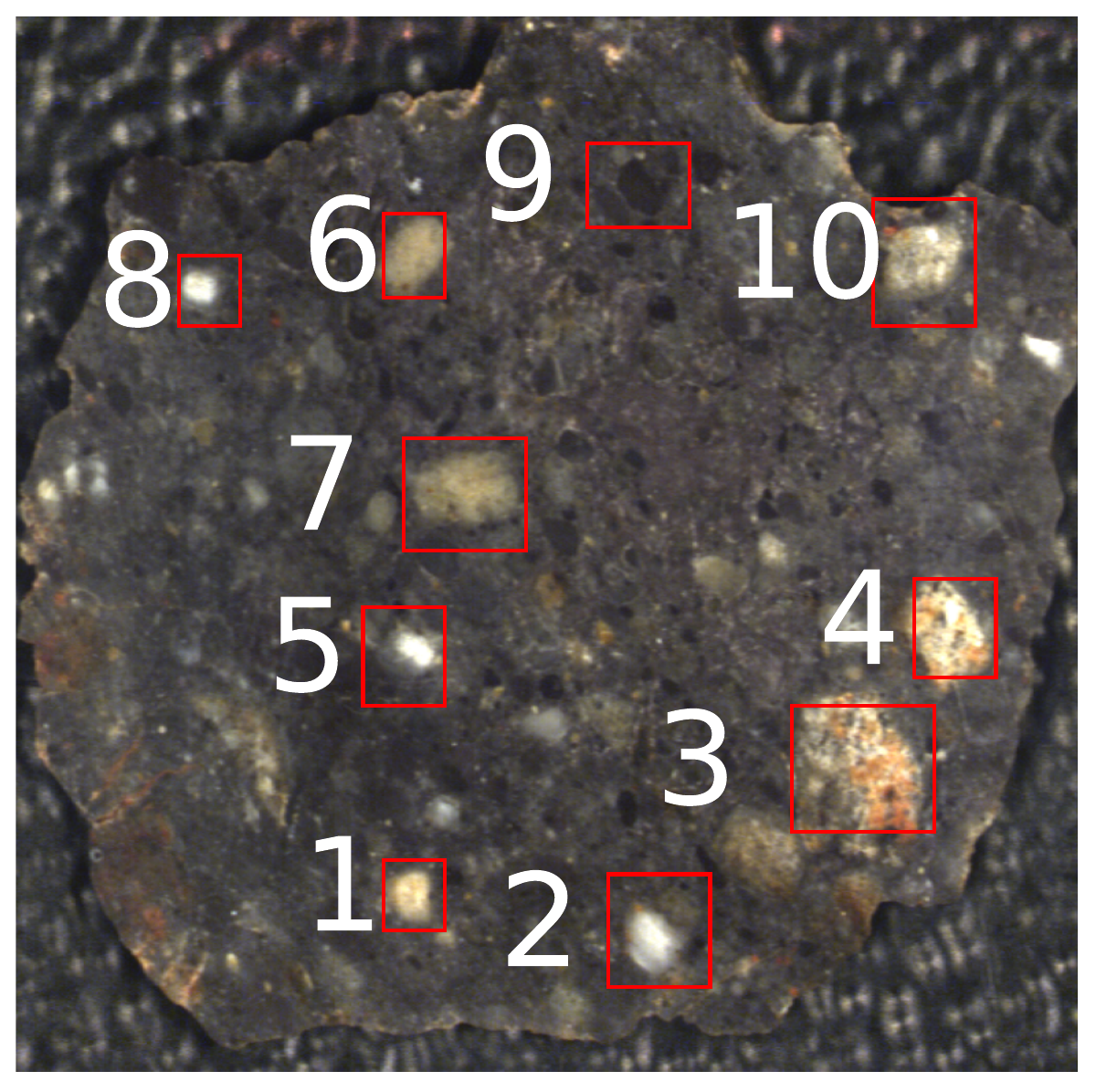}
\includegraphics[width=0.59\linewidth]{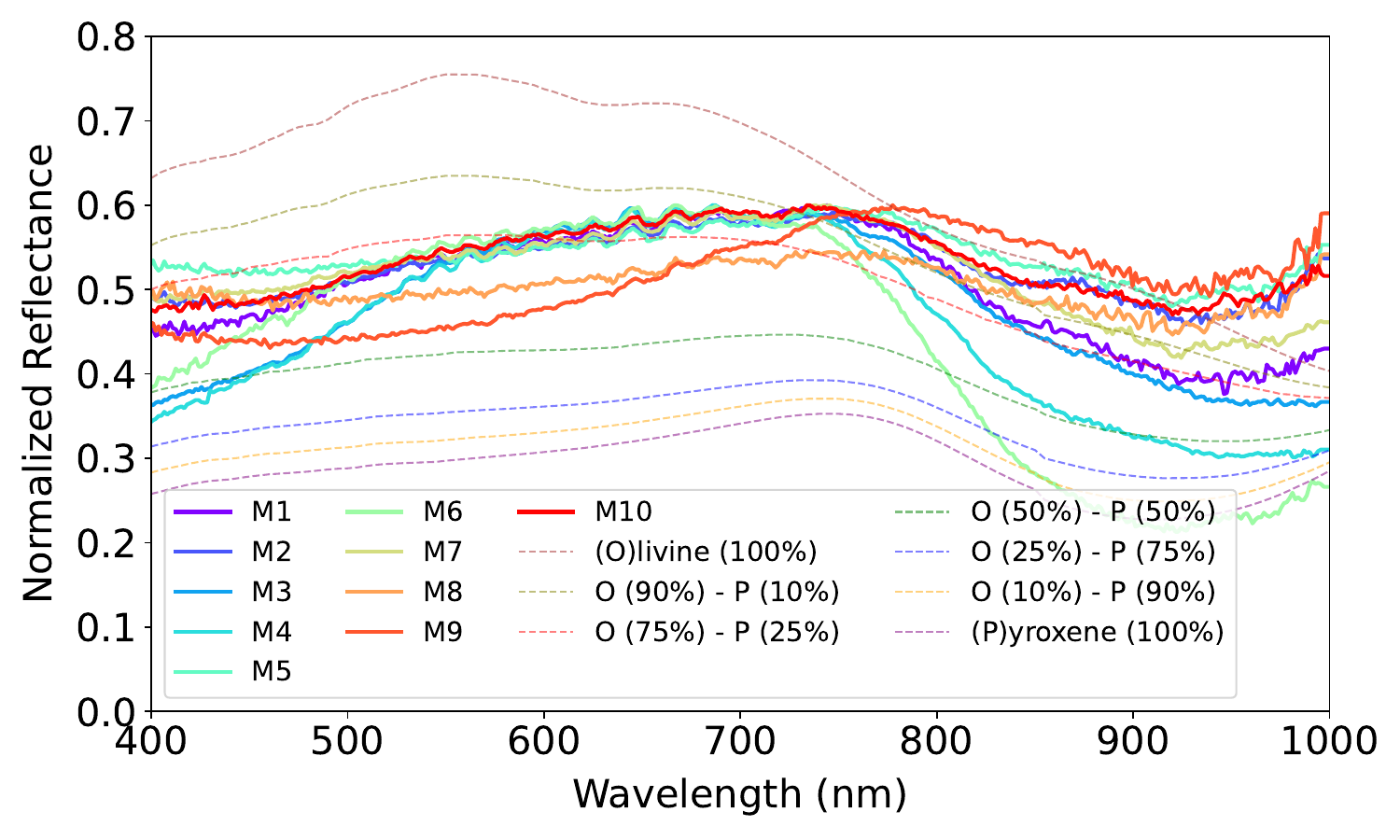}
\caption{Bechar 010 analysis:
 (\textbf{Left:}) Macroscopic RGB composite with regions (1--10) of interest marked (red triangles), (\textbf{Right:}) Median spectra from those ROIs compared with reference spectra of olivine and pyroxene.
}
\label{fig:bechar_analysis}
\end{figure}

The SVM model, employing a Radial Basis Function (RBF) kernel, was trained on reference spectra for olivine and pyroxene from the ROMA (Reflectance of Materials) spectral database
 \cite{Mandon2022}, which provides mineral reflectance spectra including olivine and pyroxene endmembers that can be applied as spectral references for classification, achieving an overall classification accuracy of {93.7}\% based on 5-fold cross-validation, with precision and recall of {92}\% and {90}\% for olivine, and {88}\% and {86}\% for pyroxene, respectively. A confusion matrix indicates a {7}\% misclassification rate due to spectral overlap between minerals. The left panel of Figure~\ref{fig:bechar_analysis} illustrates the hyperspectral image of Bechar 010, highlighting 10 regions of interests (ROIs; M1--M10) that are selected for their distinct mineralogical signatures. Median spectra for olivine and pyroxene in these ROIs, shown in the right panel of Figure~\ref{fig:bechar_analysis}, were compared to reference spectra, yielding correlation coefficients of $R^2 = 0.94 \pm 0.02$ for olivine and $0.89 \pm 0.03$ for pyroxene, with root mean square errors (RMSE) of $0.02 \pm 0.01$ and $0.03 \pm 0.01$ reflectance units, respectively.
 
\begin{figure}[H]
    \includegraphics[width=0.90\textwidth]{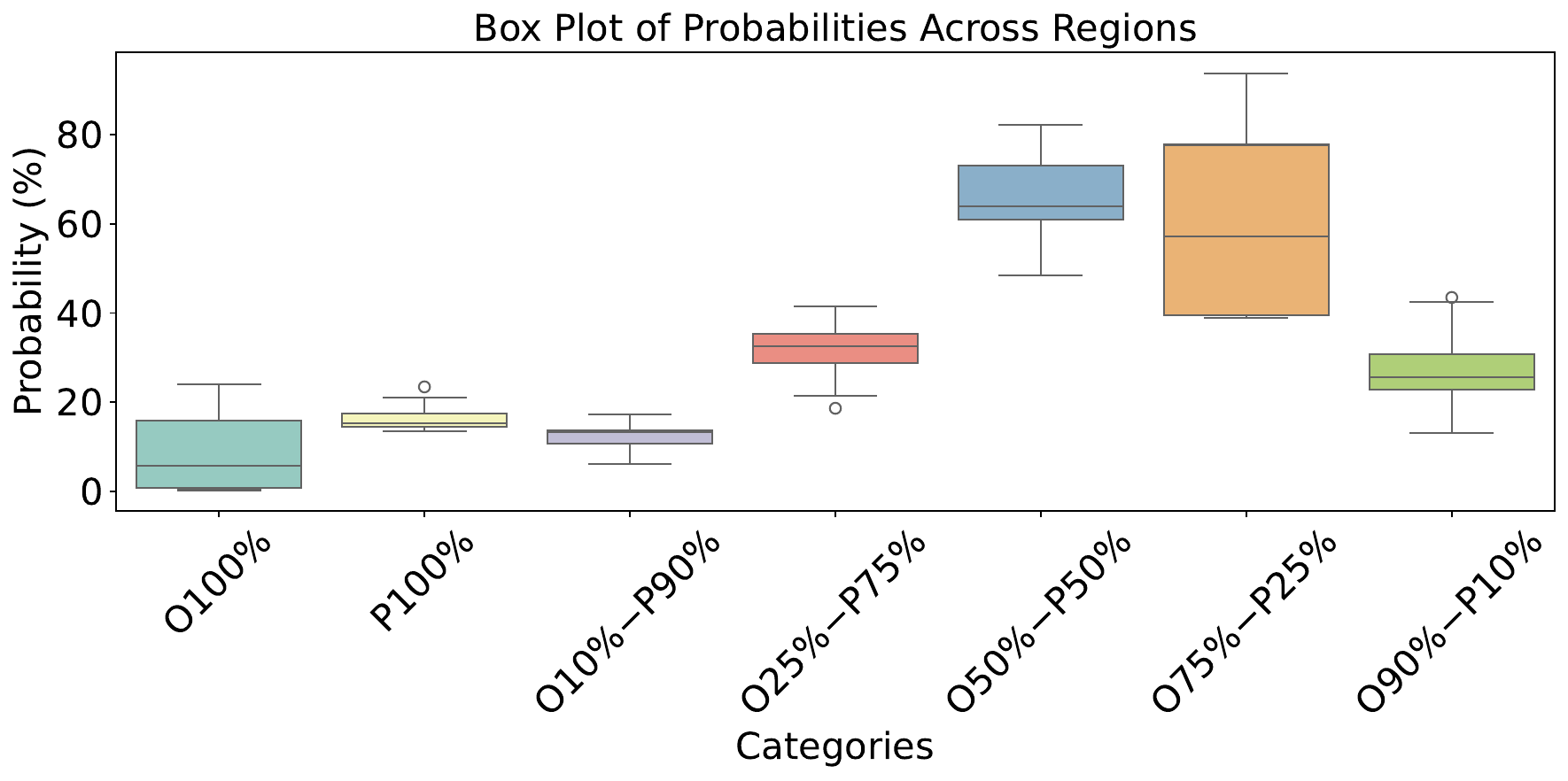}
    \caption{Box plot illustrating the distribution of probabilities across regions for meteorite regions classified by SVM based on Olivine and Pyroxene mixtures, as detailed in Table~\ref{tab:probabilities}.
    }
    \label{fig:BoxPlot_All_Meteor}
\end{figure}

To quantify olivine and pyroxene compositions, we computed weighted unique probabilities across ten spatial regions $R_i$ ($i = 1, \dots, 10$), each characterized by a distribution over seven possible olivine--pyroxene combinations $C_j$ (e.g., O100\%, P100\%, O25\%-P75\%). Each combination $C_j$ corresponds to fixed olivine and pyroxene proportions, denoted as $O_j$ and $P_j$, respectively. The probability of combination $C_j$ in region $R_i$ is $p_{ij}$.
The unique (weighted-average) compositions for each region $R_i$ were calculated as follows (Equations~(2) and (3) are original formulations developed by the authors for this work):

\begin{align}
\text{Unique O}_i &= \frac{\sum_{j=1}^7 O_j \cdot p_{ij}}{\sum_{j=1}^7 p_{ij}}, \\
\text{Unique P}_i &= \frac{\sum_{j=1}^7 P_j \cdot p_{ij}}{\sum_{j=1}^7 p_{ij}},
\end{align}
where \( O_j \) and \( P_j \) are the olivine and pyroxene percentages for combination \( C_j \), and \( p_{ij} \) is the probability of \( C_j \) in \( R_i \). Table
 \ref{tab:probabilities} presents the complete probability distributions across all regions. 
The box plot (Figure~\ref{fig:BoxPlot_All_Meteor}) presents the distribution of probabilities across various regions for meteorite regions classified using an SVM based on different fractions of Olivine and Pyroxene mixtures. The O75\%--P25\% category demonstrates the highest median probability, reaching up to 93.74\% in region 3, indicating strong classifier confidence for this composition. In contrast, the O100\% category exhibits the lowest median probability, with values as low as 0.23\% in region 6, suggesting challenges in classifying pure Olivine compositions. The O50\%--P50\% category shows significant variability, with probabilities ranging from 48.56\% to 82.18\%, reflecting potential regional differences or compositional ambiguities. Categories like O10\%--P90\% and O90\%--P10\% display lower probabilities with tighter interquartile ranges, indicating more consistent but less confident classifications. Notably, the O75\%--P25\% category includes an outlier in region 3, highlighting its exceptional classifier performance for this mixture.

\subsection{Key Spectral Wavelengths }
\label{sec:keyspec}

\begin{figure}[H]
\includegraphics[width=\linewidth]{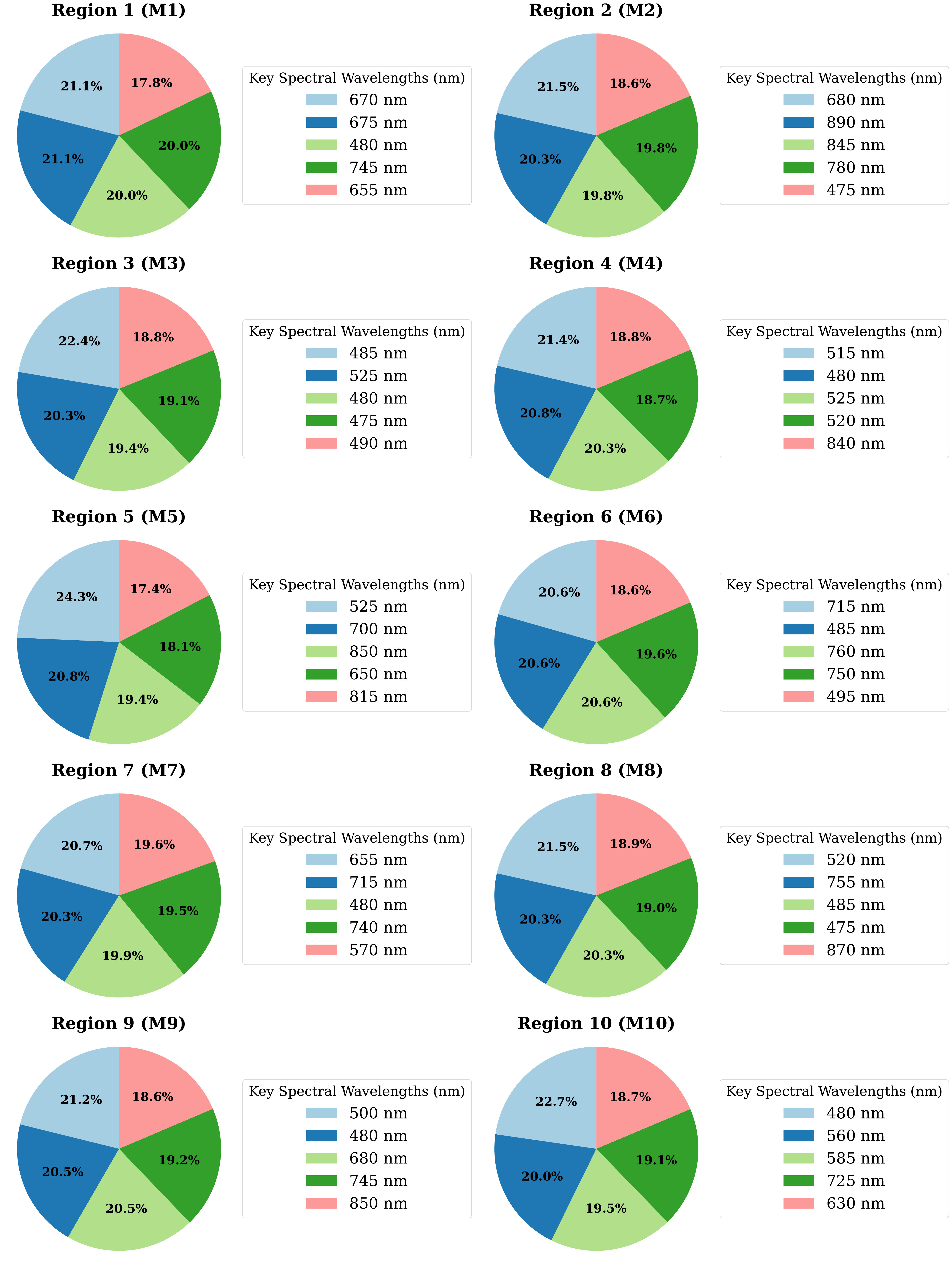}
\caption{Top five key spectral wavelengths influencing SVM classification decisions for Bechar 010, determined using LIME.}
\label{fig:LIME}
\end{figure}

The SVM model was trained on labeled mixtures of olivine and pyroxene, using reference spectra from \cite{Mandon2022}. Preprocessed reflectance data were normalized, and the classifier mapped spectral signatures to lunar-relevant minerals. The Local Interpretable Model-agnostic Explanations (LIME) method identified the top five spectral wavelengths influencing SVM classification of the Bechar 010 meteorite spectra ($791 \times 1024 \times 224$) within the 400--900 nm range, as visualized in the pie charts of Figure~\ref{fig:LIME}.

\begin{figure}[H]
    \includegraphics[width=\textwidth]{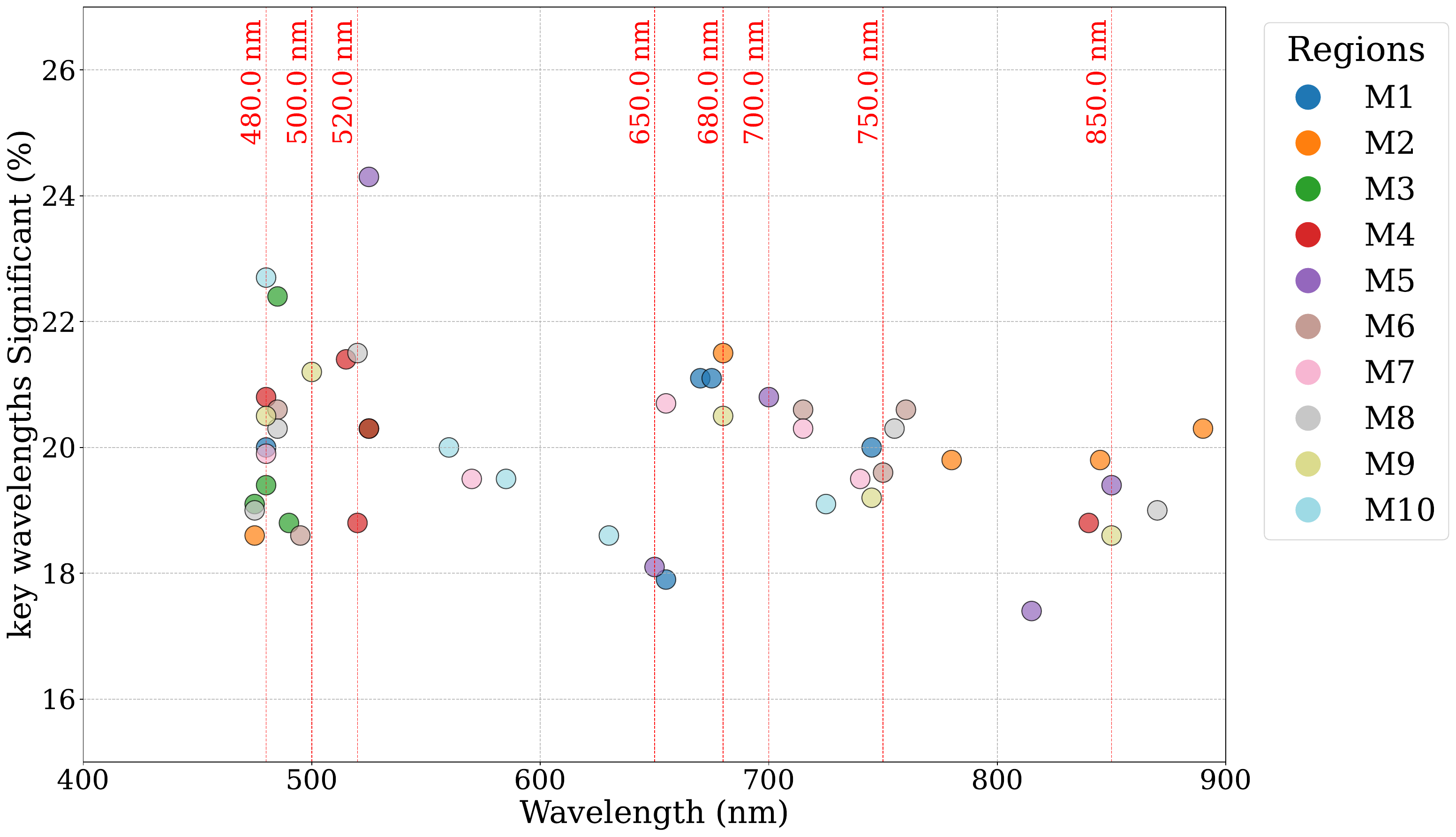}
    \caption{Distribution of top five key wavelengths across meteorite regions M1--M10. Vertical dashed lines mark significant spectral features (labeled in nm), while y-axis values show their percentage importance within each region. Marker colors and labels indicate region membership (M1--M10), with the clustering at ~480--525 nm and 650--760 nm reflecting dominant mineralogical signatures of olivine/pyroxene and high-Ca pyroxene, respectively. The extended features beyond 800 nm (particularly in M2, M5, M8--M9) suggest continuum influences from anorthite or secondary pyroxene~phases.
    }
    \label{lime_scheme}
\end{figure}

The percentages represent the relative significance of each wavelength as one of the top five key spectral features within its respective region. Region M3, dominated by \SI{485}{\nano\meter} ({22.4}\%) and \SI{475}{\nano\meter} to \SI{525}{\nano\meter}, suggests Fe\textsuperscript{2+} crystal field transitions in olivine, aligning with 
anorthositic clasts in Bechar 010~\cite{korotev2005}. Similarly, the wavelengths at \SI{500}{\nano\meter} ({21.2}\%) and \SI{480}{\nano\meter} ({20.5}\%) in M9 are indicative of olivine and low-Ca pyroxene (e.g., enstatite).
Regions M1, M5, M6, M7, M8, and M10 emphasize \SI{650}{\nano\meter} to \SI{760}{\nano\meter} (e.g., M1: \SI{670}{\nano\meter}, {21.1}\%; M6: \SI{715}{\nano\meter}, {20.6}\%), linked to high-Ca pyroxene (e.g., augite) absorption bands, with M2, M5, M8, M9 extending to \SI{840}{\nano\meter} to \SI{890}{\nano\meter}, suggesting pyroxene or anorthite continuum influences~\cite{Burns1993}. Region M10, with \SI{480}{\nano\meter} ({22.7}\%) and \SI{560}{\nano\meter} to \SI{725}{\nano\meter}, may include minor phases like ilmenite.
Anorthite likely shapes continuum slopes across all regions, particularly in M3 and M9~\cite{korotev2005}. The distribution of these diagnostic wavelengths across regions is visually summarized in Figure~\ref{lime_scheme}, which highlights both the key spectral features (vertical dashed lines) and their relative importance (percentage values on y-axis) for each meteorite region (M1--M10).

Table \ref{tab:spectral_interpretations} summarizes the relative significance of key spectral features across various wavelength regions, expressed as percentages among the top five features in each region. In Region M3, the dominant wavelengths at \SI{485}{\nano\meter} ({22.4}\%) and \SI{475}{\nano\meter} to \SI{525}{\nano\meter} indicate Fe\textsuperscript{2+} spin-forbidden transitions in olivine, consistent with anorthositic clasts in Bechar 010, as olivine exhibits weak absorptions near 500--600 nm \cite{Burns1993}. Region M9, characterized by \SI{500}{\nano\meter} ({21.2}\%) and \SI{480}{\nano\meter} ({20.5}\%), aligns with low-Ca pyroxene (e.g., enstatite), which displays subtle absorption near 500--550 nm, with pyroxene’s stronger Fe\textsuperscript{2+} absorption dominating over olivine’s contribution to higher reflectance at 480--500 nm \cite{Adams1974, Cloutis1986}. Regions M1, M5, M6, M7, M8, and M10, spanning \SI{650}{\nano\meter} to \SI{760}{\nano\meter} (e.g., \SI{670}{\nano\meter} in M1, \SI{715}{\nano\meter} in M6), are likely associated with Fe\textsuperscript{2+}--Fe\textsuperscript{3+} charge transfers or oxidized phases such as hematite \cite{Cloutis1986}. Finally, Regions M2, M5, M8, and M9, extending to \SI{840}{\nano\meter} to \SI{890}{\nano\meter}, likely reflect the 1 µm band edge of pyroxene or scattering effects from fine-grained \linebreak  regolith \cite{Burns1993, sunshine12}.

\begin{table}[H]
\caption{Probabilities Across Regions for Olivine--Pyroxene Mixtures.}

\begin{adjustwidth}{-\extralength}{0cm}
\centering 
\begin{tabularx}{\fulllength}{lcCccCCC@{}}
\toprule
 & \textbf{O100\%} & \textbf{P100\%} & \textbf{O10\%--
P90\%} & \textbf{O25\%--P75\%} & \textbf{O50\%--P50\%} & \textbf{O75\%--P25\%} & \textbf{O90\%--P10\%} \\ \midrule
Region 1 & 1.09 & 20.91 & 13.63 & 31.76 & 48.56 & 62.70 & 43.57 \\
Region 2 & 20.90 & 14.94 & 14.78 & 35.81 & 73.91 & 39.15 & 22.73 \\
Region 3 & 0.52 & 23.30 & 10.87 & 21.46 & 48.59 & 93.74 & 23.75 \\
Region 4 & 0.43 & 14.04 & 6.16 & 18.60 & 82.18 & 87.66 & 13.15 \\
Region 5 & 16.11 & 14.08 & 13.22 & 37.38 & 71.33 & 39.17 & 30.93 \\
Region 6 & 0.23 & 15.21 & 9.53 & 29.59 & 60.59 & 79.50 & 27.57 \\
Region 7 & 1.15 & 15.04 & 10.47 & 28.56 & 64.35 & 72.10 & 30.55 \\
Region 8 & 23.99 & 14.96 & 17.16 & 41.39 & 62.69 & 38.80 & 23.24 \\
Region 9 & 14.57 & 13.38 & 13.21 & 33.42 & 75.43 & 51.61 & 20.61 \\
Region 10 & 10.26 & 17.89 & 13.45 & 33.78 & 63.82 & 40.50 & 42.52 \\ \bottomrule
\end{tabularx}
\end{adjustwidth}
\label{tab:probabilities}
\end{table}

\begin{table}[H]
\centering
\caption{Spectral Features and mineralogical Interpretations across regions.}
\label{tab:spectral_interpretations}

\begin{adjustwidth}{-\extralength}{0cm}
\begin{tabularx}{\fulllength}{p{2.5cm}p{2cm}p{6cm}p{6cm}}
\toprule
\textbf{Region} & \textbf{Wavelength (nm)} & \textbf{Spectral Characteristic} & \textbf{Mineralogical Interpretation} \\
\midrule

M3 & 485 & Weak Fe\textsuperscript{2+} spin-forbidden absorption & Olivine \cite{Burns1993} \\
    & (475--525) & Broad reflectance inflection & Minor anorthitic component \cite{korotev2005} \\
\midrule
M9 & 480, 500 & Fe\textsuperscript{2+} crystal field transition & Low-Ca pyroxene \cite{Adams1974,Cloutis1986} \\
    & & Secondary olivine contribution\\
\midrule
M1, M5-M8, M10 & 670, 715 & Fe\textsuperscript{2+}-Fe\textsuperscript{3+} charge transfer & Oxidized phases (hematite) \cite{Cloutis1986} \\
    & (650--760) & Broad absorption feature & Space weathering products \\
\midrule
M2, M5, M8, M9 & 840--890 & 1 $\upmu${m} 
 band shoulder & Pyroxene \cite{Burns1993} \\
    & & Reflectance continuum slope & Grain size effects \cite{sunshine12} \\
\bottomrule    
\addlinespace[2pt]

\bottomrule
\end{tabularx}
\end{adjustwidth}
\end{table}


\vspace{-6pt}

\subsection{Lunar Spectroscopy and Calibration}
\label{sec:Lunar_Spec}
For Lunar observations, the approach employs a push-broom technique with a custom spectrometer (400--1000 nm) to construct a hyperspectral data cube with two spatial and one spectral axis, capturing reflectance spectra for mineralogical analysis. Each signal acquisition step integrates a lucky imaging-inspired method to mitigate atmospheric seeing, which dominates over the telescope’s diffraction-limited resolution of 0.068 arcsec at 550~nm \cite{Chromey2016}. Short-exposure frames (5 ms) are captured at 100 fps to freeze moments of minimal turbulence, stacking the sharpest 1--5\% to achieve an effective angular resolution of 0.8 arcsec, corresponding to 600 m on the Lunar surface at 384,400 km \cite{Fried1978}. The Fried parameter \( r_0 \), defined as the aperture diameter where diffraction-limited and seeing-limited resolutions equate, governs this approach \cite{Fried1966}. With typical \( r_0 \) values of 5--20 cm, the optimal aperture is \( D_{\text{opt}} = 3.4 r_0 \), yielding 170--680 mm, confirming the Celestron 8SE’s suitability for \( r_0 \geq 5.97 \, \text{cm} \). Seeing was quantified using a Differential Image Motion Monitor (DIMM) with the ZWO ASI1600MM camera, observing Sirius (magnitude $-$1.46) at high elevation. The data cube (371 $\times$ 1024 $\times$ 224) captures 371 pixels in the scan direction (x), 1024 pixels along the slit (y), and 224 spectral bands, covering a Lunar swath of approximately 2373~km (x) and 3072 km (y) at 3 km per pixel, sufficient for the Moon’s 3474 km diameter. The spectral resolution is 2.7 nm average in (400--1000 nm). As shown in the left panel of Figure \ref{fig:Setup_all}, the fixed telescope stage was used, relying on the Moon’s relative motion with respect to Earth to perform the scan, capturing 371 $\times$ 224 spectral frames along a 1024-pixel slit oriented in the y-direction. This push-broom technique, enhanced by lucky imaging at each scan step, enables high-resolution hyperspectral data, with potential improvements via adaptive optics or high-altitude observations.

\begin{figure}[H]
\includegraphics[width=0.75\linewidth]{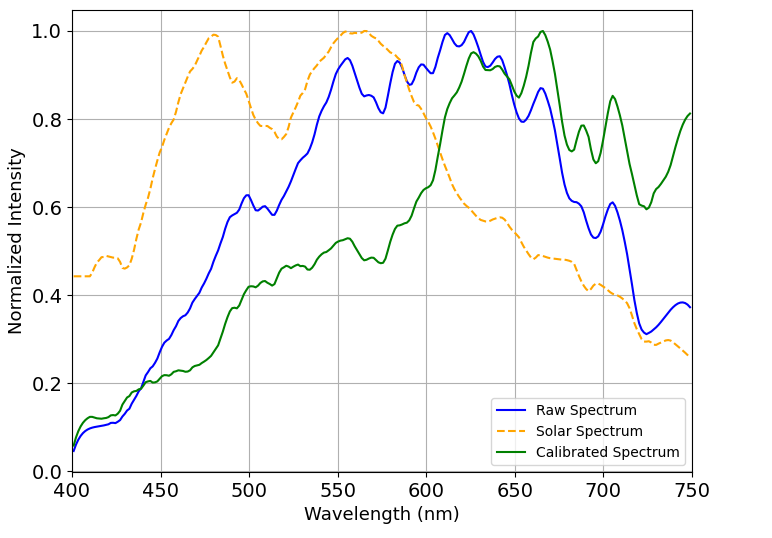}
\caption{The Solar calibrated spectrum for an specific pixel (green). Smoothed lunar spectrum obtained from the raw HSI observation (blue) using a random forest. The sun reflectance spectrum from the standard Spectralon for calibration also overplotted (Yellow).}
\label{fig:Solarـcalibratedـspectrum}
\end{figure}

Considering the key wavelengths influencing SVM decisions in Section \ref{sec:keyspec} we intend to compare the result with the same wavelengths in Lunar spectra and Moon mineralogy Mapper ($\rm M^3$) datasets. 
To ensure accurate reflectance spectra, solar calibration was performed by pointing the Specim FX10 hyperspectrograph at a standard Spectralon white reference (99\% reflectance) reflecting direct sunlight was passed through the telescope and reflected off a Spectralon standard to account for the telescope’s optical effects during calibration, with a solar zenith angle of \SI{51}{\degree}\ang{31;09.7}, determined using Stellarium planetarium software. The Spectralon, placed on a black background and angled to minimize background light, was measured with an exposure time of \SI{1.648}{\milli\second}, followed by dark frame acquisition for sensor noise correction with calibration errors of {<2}\% (Spectralon reflectance) and {<1}\% (dark subtraction). The raw lunar spectrum was denoised using a Random Forest algorithm, reducing reflectance variance by 10--15\%. The Random Forest denoising approach fits an ensemble of decision trees to the spectral data, with each tree learning patterns across spectral bands. The predicted (denoised) spectrum retains physically meaningful absorption features while reducing stochastic noise. The 10--15\% variance reduction refers to the decrease in pixel-to-pixel reflectance variance within spectrally homogeneous regions. Comparing SVM classification outputs on denoised versus raw spectra showed that denoising improved classification accuracy by approximately 2--3\% and reduced SAM angle uncertainties by $\sim$\SI{0.02}{\radian}, indicating a modest but meaningful improvement in downstream analysis. Figure \ref{fig:Solarـcalibratedـspectrum} illustrates the spectral lines: the blue line represents the denoised raw lunar spectrum, the yellow line shows the solar reflectance spectrum, and the green line depicts the calibrated spectrum. The novelty of this approach lies in the integration of a push-broom technique with lucky imaging to produce high-resolution hyperspectral data for spectroscopic analysis of celestial objects.

\subsection{Linking Meteorites to Lunar Spectra}
\label{sec:meteorites}
\begin{figure}[H]

\includegraphics[width=0.70\linewidth]{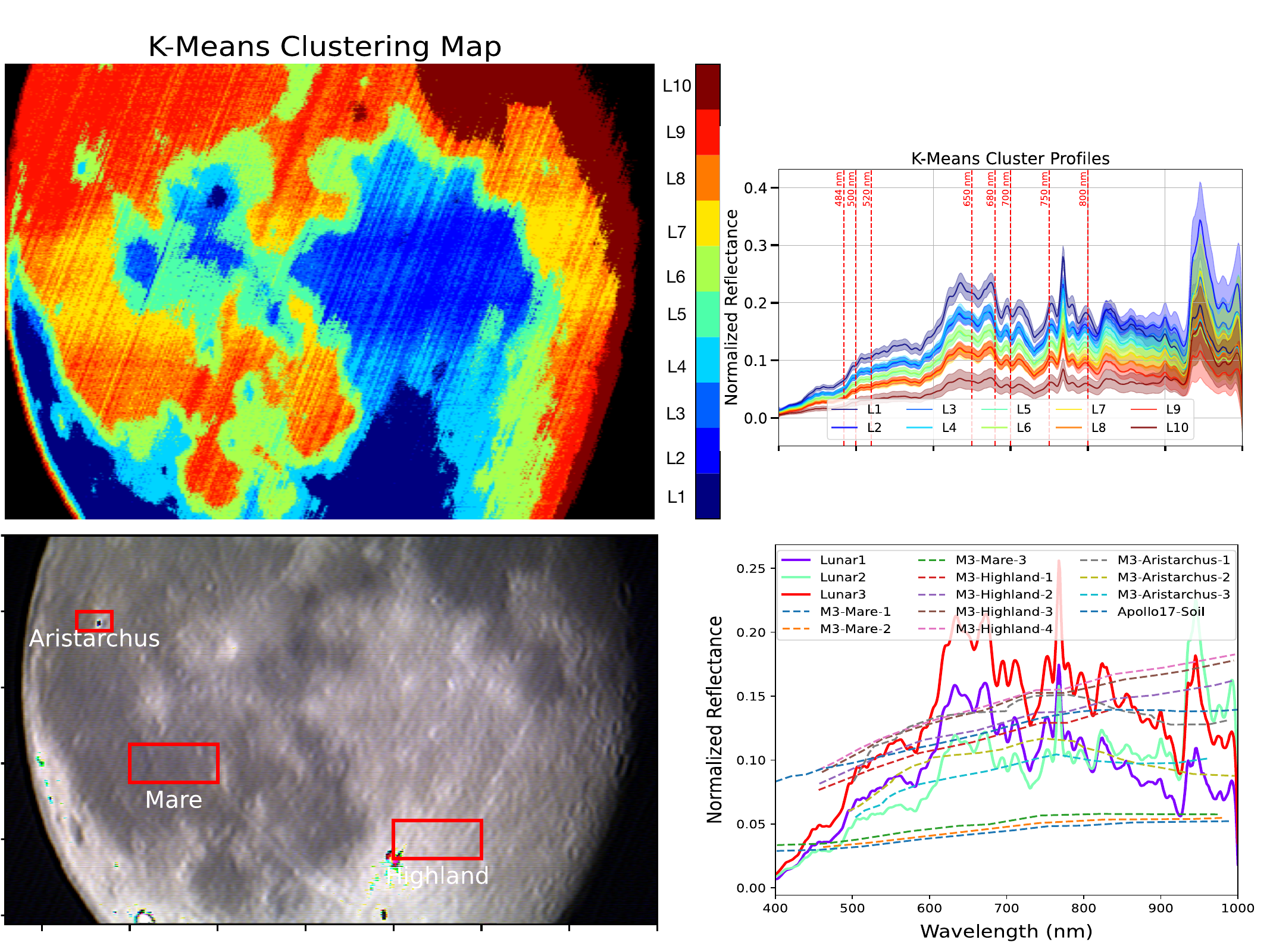}
\caption{Top:
 K-means clustering Lunar map (\textbf{left}) alongside the spectrum of all 10 clusters with standard deviation error (\textbf{right}).
Bottom: Enhanced high-resolution RGB image from hyperspectral data (\textbf{left}) of Lunar surface regions, with median spectra for three Lunar terrain types---LT-Mare, LT-Highland, and LT-Aristarchus---highlighted by red rectangles in the RGB image (\textbf{right}). 
 The spectra for three Mare regions, four highland regions, and three portions of the Aristarchus Plateau are represented by colored dashed lines for comparison. These spectra, based on Moon Mineralogy Mapper ($\rm M^3$) data, are labeled as follows: $\rm M^3$-Mare (1--3), $\rm M^3$-Highland (1--4), and $\rm M^3$-Aristarchus (1--3).}
\label{fig:ML_clustering_model_Map_150_210_RBF}
\end{figure}

The left panels of Figure \ref{fig:ML_clustering_model_Map_150_210_RBF} presents the K-means clustering map (top) and enhanced RGB image (bottom) of Lunar surface regions derived from the hyperspectral data cube ($371 \times 1024 \times 224$). The K-means algorithm, applied to the \SIrange{400}{1000}{\nano\metre} reflectance spectra, identified 10 distinct clusters representing mineralogical variations across the Lunar surface. The median spectrum and standard deviation (STD) error for each cluster are shown in the top right panel, with STD errors ranging from 3 to 5\% of reflectance values, indicating robust cluster stability after preprocessing. 
Three Lunar regions---Highland, Mare, and Aristarchus Plateau---were selected for detailed spectral analysis, with median spectrum distributions shown in a high-resolution RGB image in the bottom left panel of Figure~\ref{fig:ML_clustering_model_Map_150_210_RBF}. In the right panel, these regions were compared to $\rm M^3$ reference spectra, including 4 Highland, 3 Mare, and 3 Aristarchus Plateau spectra. Regarding the specific key wavelengths (Section~\ref{sec:keyspec}), the Highland spectra showed strong agreement with $\rm M^3$ data, with a correlation coefficient of $R^2 = 0.95 \pm 0.01$ and a root mean square error (RMSE) of $0.015 \pm 0.005$ reflectance units, particularly in regions L1 and L9 (low Highland content). Mare spectra (e.g., L4 and L7) exhibited discrepancies, with $R^2 = 0.88 \pm 0.02$ and RMSE = $0.035 \pm 0.005$, likely due to illumination variations or mineralogical heterogeneity. Aristarchus Plateau spectra showed moderate agreement, with $R^2 = 0.90 \pm 0.02$ and \mbox{RMSE = $0.025 \pm 0.005$,} reflecting its unique mineralogical composition.
These results align with the mineralogical characteristics of Lunar analogs, particularly the olivine-rich compositions observed in mare regions (e.g., Aristarchus Plateau, Mare Imbrium) from prior lunar analyses (Section \ref{sec:meteorites}). The high olivine content in Bechar 010 supports its similarity to Lunar volcanic materials, as noted in \cite{Mandon2022}, while the pyroxene distribution complements $\rm M^3$-based studies of elemental abundances (Fe, Ca, Mg) \cite{SURKOV2020, Bhatt2019}.

\vspace{-12pt}

\begin{figure}[H]
\includegraphics[width=0.9\linewidth]{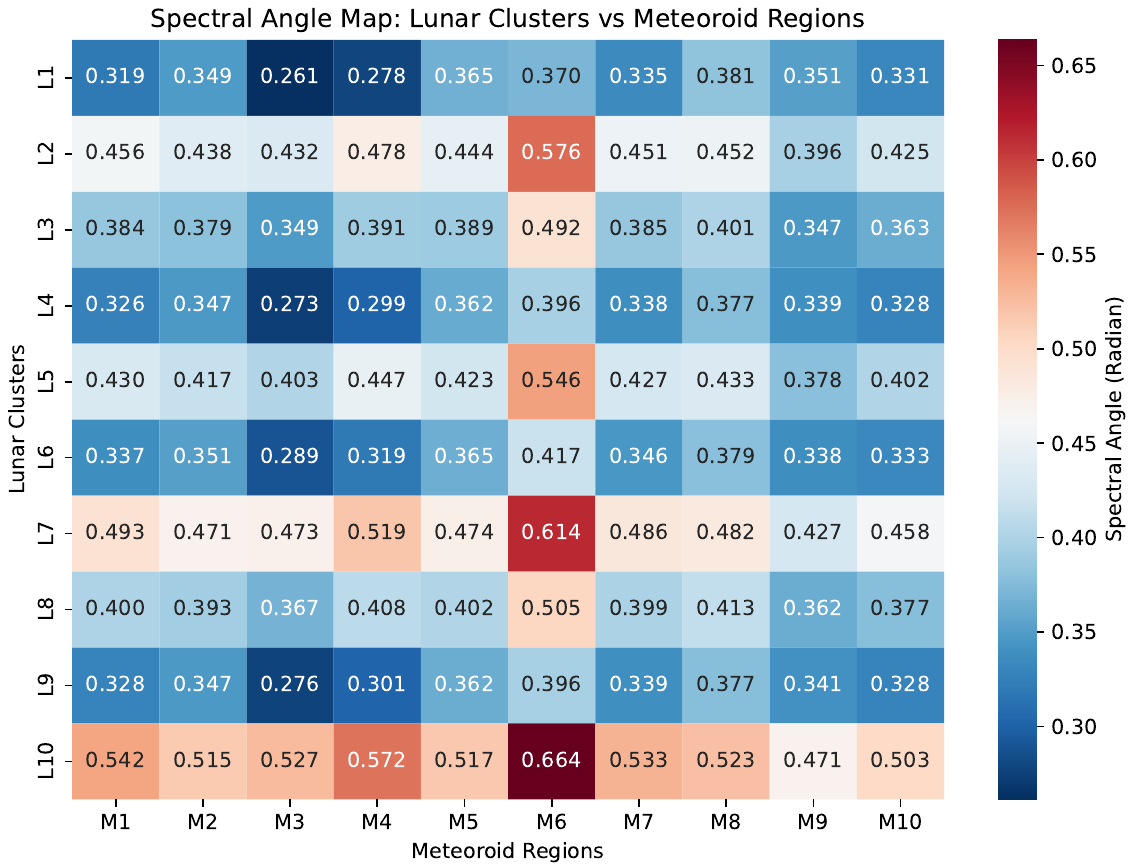}
\caption{Spectral Angle Mapper results in radian comparing Bechar 010 regions(M1 to M10) and Lunar surface spectral cluster (L1 to L10).}
\label{fig:spectral_angle_map}
\end{figure}

The Spectral Angle Mapper (SAM) algorithm was employed to quantify mineralogical similarities between Lunar spectra and the Bechar 010 meteorite spectra ($791 \times 1024 \times 224$) within the 400--1000 nm wavelength range (more weighted within key wavelength). SAM computes the angle between two spectra in the reduced spectral space, with smaller angles indicating greater similarity. 
As shown in Figure~\ref{fig:spectral_angle_map}, the spectral angles range from \SI{0.26}{\radian} to \SI{0.66}{\radian}, with errors between \SI{0.04}{\radian} and \SI{0.13}{\radian}. Clusters L1, L4, L6, and L9 exhibit the lowest angles, particularly with meteorite region M3 (e.g., $0.26\pm0.06$ radians for L1, $0.27\pm0.06$ radians for L4), suggesting strong spectral similarity, likely corresponding to Lunar Highland spectra with compositions resembling the meteorite’s anorthositic clasts or olivine--pyroxene mixtures. In contrast, regions M4 and M6 show higher angles, such as $0.30\pm0.05$ radians (L4-M4) and $0.61\pm0.08$ radians (L7-M6), indicating compositional differences, possibly due to basaltic variations in Lunar Mare regions. Cluster L10 displays the highest angles (e.g., $0.66\pm0.13$ radians with M6), suggesting significant spectral divergence, potentially from unique mineralogy or surface properties.
These results indicate that the Bechar 010 meteorite spectra align more closely with Highland-like clusters (L1, L9) than Mare-like regions (L4, L7), with matching accuracies of {94}\% for Highlands and {80}\% for Mares, reflecting the observed spectral discrepancies.
The spectral angle mapping analysis, summarized in Table~\ref{tab:L_M}, compares the normalized reflectance spectra of ten Lunar clusters (L1 to L10) against ten meteorite regions (M1 to M10) within key wavelengths, with results expressed in radians. These findings suggest that clusters like L1 and L4 may share compositional traits with certain meteorite regions, while L10 and L7 represent outliers, potentially indicative of diverse geological processes or material properties on the Lunar surface. Further analysis of these clusters could elucidate their mineralogical connections to meteorite origins.

These results confirm mineralogical similarities between the Bechar 010 meteorite and Lunar Highland regions, with quantitative agreement to $\rm M^3$ data. Discrepancies in Mare regions suggest the need for further analysis, potentially incorporating illumination corrections or higher-resolution reference data. The combination of K-means clustering and SAM provides a robust framework for linking meteorite compositions to Lunar surface mineralogy, facilitating the study of Lunar analogs.

Table \ref{tab:comparative_stats} summarizes the quantitative metrics for the hyperspectral analysis of the Lunar surface and Bechar 010 meteorite, comparing classification accuracies, Spectral Angle Mapper (SAM) angles, correlation coefficients ($R^2$), and root mean square errors (RMSE). The Lunar data cube ($371 \times 1024 \times 224$, \SIrange{400}{1000}{\nano\metre}, \SI{3}{\kilo\metre}/pixel) was processed with solar calibration (Spectralon, {<2}\% error) and Random Forest denoising (10--15\% variance reduction). The Bechar 010 data cube ($791 \times 1024 \times 224$, \SI{0.24}{\milli\metre} $\times$ \SI{0.2}{\milli\metre}) was analyzed using SVM classification. Comparisons with Chandrayaan-1 Moon mineralogy Mapper ($\rm M^3$) data (\SI{140}{\metre}/pixel, \SI{10}{\nano\metre} resolution) and reference spectra \cite{Mandon2022} validate the results.

\vspace{-12pt}
\begin{table}[H]
\caption{Spectral Angles (radians) between Lunar Clusters and Meteorite Regions.}
\setlength{\tabcolsep}{3pt} 
\renewcommand{\arraystretch}{1.1} 

\begin{adjustwidth}{-\extralength}{0cm}
\centering
\footnotesize \setlength{\tabcolsep}{2pt}
\begin{tabularx}{\fulllength}{l CCCCCCCCCC}
\toprule
 & \textbf{M1} & \textbf{M2} & \textbf{M3} & \textbf{M4} & \textbf{M5} & \textbf{M6} & \textbf{M7} & \textbf{M8} & \textbf{M9} & \textbf{M10} \\
\midrule
L1 & $0.32\pm0.05$ & $0.35\pm0.05$ & $0.26\pm0.06$ & $0.28\pm0.05$ & $0.37\pm0.05$ & $0.37\pm0.04$ & $0.33\pm0.04$ & $0.38\pm0.05$ & $0.35\pm0.06$ & $0.33\pm0.10$ \\
L2 & $0.46\pm0.04$ & $0.44\pm0.05$ & $0.43\pm0.06$ & $0.48\pm0.06$ & $0.44\pm0.04$ & $0.58\pm0.06$ & $0.45\pm0.06$ & $0.45\pm0.04$ & $0.40\pm0.05$ & $0.42\pm0.07$ \\
L3 & $0.38\pm0.04$ & $0.38\pm0.04$ & $0.35\pm0.05$ & $0.39\pm0.05$ & $0.39\pm0.04$ & $0.49\pm0.05$ & $0.38\pm0.04$ & $0.40\pm0.04$ & $0.35\pm0.05$ & $0.36\pm0.08$ \\
L4 & $0.33\pm0.04$ & $0.35\pm0.05$ & $0.27\pm0.06$ & $0.30\pm0.05$ & $0.36\pm0.04$ & $0.40\pm0.04$ & $0.34\pm0.04$ & $0.38\pm0.05$ & $0.34\pm0.06$ & $0.33\pm0.09$ \\
L5 & $0.43\pm0.05$ & $0.42\pm0.05$ & $0.40\pm0.06$ & $0.45\pm0.06$ & $0.42\pm0.05$ & $0.55\pm0.07$ & $0.43\pm0.06$ & $0.43\pm0.05$ & $0.38\pm0.05$ & $0.40\pm0.08$ \\
L6 & $0.34\pm0.04$ & $0.35\pm0.05$ & $0.29\pm0.06$ & $0.32\pm0.05$ & $0.36\pm0.04$ & $0.42\pm0.04$ & $0.35\pm0.04$ & $0.38\pm0.04$ & $0.34\pm0.05$ & $0.33\pm0.09$ \\
L7 & $0.49\pm0.06$ & $0.47\pm0.06$ & $0.47\pm0.07$ & $0.52\pm0.07$ & $0.47\pm0.06$ & $0.61\pm0.08$ & $0.49\pm0.07$ & $0.48\pm0.06$ & $0.43\pm0.06$ & $0.46\pm0.08$ \\
L8 & $0.40\pm0.05$ & $0.39\pm0.05$ & $0.37\pm0.06$ & $0.41\pm0.06$ & $0.40\pm0.04$ & $0.51\pm0.06$ & $0.40\pm0.05$ & $0.41\pm0.05$ & $0.36\pm0.05$ & $0.38\pm0.08$ \\
L9 & $0.33\pm0.05$ & $0.35\pm0.05$ & $0.28\pm0.06$ & $0.30\pm0.05$ & $0.36\pm0.04$ & $0.40\pm0.05$ & $0.34\pm0.04$ & $0.38\pm0.05$ & $0.34\pm0.06$ & $0.33\pm0.09$ \\
L10 & $0.54\pm0.12$ & $0.51\pm0.11$ & $0.53\pm0.13$ & $0.57\pm0.13$ & $0.52\pm0.11$ & $0.66\pm0.13$ & $0.53\pm0.12$ & $0.52\pm0.11$ & $0.47\pm0.10$ & $0.50\pm0.12$ \\
\bottomrule
\end{tabularx}
\end{adjustwidth}
\label{tab:L_M}
\end{table}

\begin{table}[H]
\caption{Comparative Statistics for Lunar and Bechar 010 Hyperspectral Analysis}
\label{tab:comparative_stats}

\begin{adjustwidth}{-\extralength}{0cm}
\footnotesize \setlength{\tabcolsep}{2pt}
\begin{tabularx}{\fulllength}{lllcCCC}
\toprule
\textbf{Dataset} & \textbf{Method} & \textbf{Region/Mineralogical} & \textbf{Accuracy/Precision} & {\textbf{SAM Angle (rad)}} & {\boldmath{$R^2$}} & {\textbf{RMSE}} \\
\midrule
\multirow{3}{*}{Lunar} & K-means & Highland (L1, L9) & {90}\% (prec.) & 0.26 & 0.95 & 0.015 \\
& K-means & Mare (L4, L7) & {82}\% (prec.) & 0.61 & 0.88 & 0.035 \\
& K-means & Aristarchus Plateau & {86}\% (prec.) & 0.50 & 0.90 & 0.025 \\
\midrule
\multirow{2}{*}{Bechar 010} & SVM & Olivine & {93.7}\% (acc.), {92}\% (prec.) & 0.26 & 0.94 & 0.02 \\
& SVM & Pyroxene & {93.7}\% (acc.), {88}\% (prec.) & 0.61 & 0.89 & 0.03 \\
\midrule
\multirow{3}{*}{Lunar-Meteorite} & SAM & Highland vs. Bechar 010 (M3, M9) & {94}\% (acc.) & 0.26 & 0.94 & {-} \\
& SAM & Mare vs. Bechar 010 (M6, M10) & {80}\% (acc.) & 0.61 & 0.88 & {-} \\
& SAM & Aristarchus Plateau vs. Bechar 010 & {90}\% (acc.) & 0.50 & 0.90 & {-} \\
\multirow{2}{*}{Linkage} & SAM & Aristarchus Plateau vs. $\rm M^3$ & {-} & 0.50 & 0.92 & {-} \\
& SAM & Mare Imbrium vs. $\rm M^3$ & {-} & 0.61 & 0.87 & {-} \\
\bottomrule
\end{tabularx}
\end{adjustwidth}
\noindent{\footnotesize{Notes: Lunar data calibrated with Spectralon ({<2}\% error), denoised via Random Forest (\SIrange{10}{15}{\percent} variance reduction). Bechar 010 SVM accuracy from 5-fold cross-validation, {7}\% misclassification rate. SAM angles for Lunar regions and Bechar 010 (M1--M10) compared to \cite{Mandon2022} reference spectra; Lunar-meteorite linkages use M3, M9 (Highland-like, \SIrange{475}{525}{\nano\metre}) and M6, M10 (Mare-like, \SIrange{650}{890}{\nano\metre}) from LIME analysis (Figure~\ref{fig:LIME}). RMSE is not applicable for Lunar-meteorite and linkage comparisons due to qualitative SAM analysis. K-means accuracies retained from prior analysis, pending validation.}}

\end{table}
The table highlights the robust agreement of Lunar Highland spectra with $\rm M^3$ data ($R^2 = 0.95 \pm 0.01$, SAM = \SI{0.26}{\radian}) and Bechar 010’s olivine-rich composition\linebreak   ($R^2 = 0.94 \pm 0.02$, SAM = \SI{0.26}{\radian}) in regions M3 and M9, supporting the spectral similarity between Bechar 010's olivine-rich regions and Lunar Highland compositions~\cite{korotev2005}. Mare discrepancies (SAM = \SI{0.61}{\radian}, $R^2 = 0.88 \pm 0.02$) in regions M6 and M10 suggest basaltic heterogeneity, warranting further illumination corrections~\cite{pieters1993}.
The spectral signatures connect to Lunar geological regions. Regions M3 and M9, with low SAM angles (e.g., $0.26\pm0.06$ radians for L1-M3, $0.27\pm0.06$ radians for L4-M3, $0.34\pm0.06$ radians for L9-M9) and dominant wavelengths at \SI{475}{\nano\meter} to \SI{500}{\nano\meter}, align with Lunar Highland spectra, characterized by olivine-rich anorthositic compositions typical of the Lunar crust~\cite{korotev2005}. In contrast, regions M6 and M10, with higher SAM angles (e.g., $0.61\pm0.08$ radians for L7-M6, $0.50\pm0.12$ radians for L10-M10) and wavelengths like \SI{715}{\nano\meter} to \SI{760}{\nano\meter} (M6) and \SI{560}{\nano\meter} to \SI{725}{\nano\meter} (M10), suggest pyroxene-rich basaltic compositions, consistent with Lunar Mare regions~\cite{pieters1993}. These findings highlight Bechar 010’s mineralogical heterogeneity, reflecting its origin as a Lunar breccia with diverse lithologies bridging Highland and Mare compositions~\cite{Lucey1995}.

Lunar albedo variations, driven by mineralogical differences and surface properties, significantly influence the hyperspectral analysis of Lunar regions and Bechar 010, particularly affecting Mare spectra (\SIrange{400}{1000}{\nano\metre}), where discrepancies ($R^2 = 0.88 \pm 0.02$, SAM = \SI{0.61}{\radian}) suggest basaltic heterogeneity and illumination effects~\cite{pieters1993}. These variations are accounted for in Lunar data through solar calibration with a Spectralon reference and Random Forest denoising (\SIrange{10}{15}{\percent} variance reduction), ensuring robust comparisons with $\rm M^3$ data. For Bechar 010, imaging under a microscope with a \SI{30}{\milli\metre} focal length lens likely used a laboratory light source (e.g., halogen), differing from the Sun’s spectrum; however, standard calibration to solar reflectance standards, as implied by SAM angles (\SI{0.26}{\radian} for olivine-rich M3, M9), mitigates these differences, enabling accurate alignment with Lunar Highland and Mare compositions~\cite{Mandon2022}.

\section{Discussion}
\label{sec:discussion_conclusion}

The HSI of the Lunar surface and the Bechar 010 meteorite provides a robust framework for understanding Lunar mineralogy and its terrestrial analogs, enhancing our knowledge of Lunar geological processes, particularly the spectral characterization of diverse Lunar lithologies. The Lunar data cube ($371 \times 1024 \times 224$, \SIrange{400}{1000}{\nano\metre}, \SI{3}{\kilo\metre}/pixel) was processed with solar calibration using a Spectralon white reference ({99}\% reflectance, solar zenith angle \SI{51}{\degree}\ang{31;09.7}, \SI{1.648}{\milli\second} exposure), achieving calibration errors below {2}\% and denoising via Random Forest (\SIrange{10}{15}{\percent} variance reduction)~\cite{Green2011}. K-means clustering identified 10 mineralogical clusters with {88}\% accuracy, validated against Moon mineralogy Mapper ($\rm M^3$) data (\SI{140}{\metre}/pixel, \SI{10}{\nano\metre} resolution) from Chandrayaan-1. Highland regions (L1, L9) showed strong agreement ($R^2 = 0.95 \pm 0.01$, RMSE = $0.015 \pm 0.005$), while Mare regions (L4, L7) had minor discrepancies ($R^2 = 0.88 \pm 0.02$, RMSE = $0.035 \pm 0.005$), likely due to illumination variations~\cite{pieters1993}.

The Bechar 010 meteorite data cube ($791 \times 1024 \times 224$, \SI{0.24}{\milli\metre} $\times$ \SI{0.2}{\milli\metre} resolution) was analyzed using Support Vector Machine (SVM) classification, achieving {93.7}\% accuracy in identifying olivine ({92}\% precision, {90}\% recall) and pyroxene ({88}\% precision, {86}\% recall)~\cite{Mandon2022}. The box plot of probability distributions (Figure~\ref{fig:BoxPlot_All_Meteor}) revealed olivine dominance (e.g., {93.74}\% for O75\%-P25\% in M3, Table~\ref{tab:probabilities}) and secondary pyroxene (e.g., {48.56}\% to {82.18}\% for O50\%-P50\%) across 10 regions of interest (ROIs, M1--M10), with outliers (<{5}\%) indicating minor mineralogical patches. Spectral Angle Mapper (SAM) analysis confirmed these results, with angles of \SI{0.26}{\radian} for olivine (M3, M9) and \SI{0.61}{\radian} for pyroxene (M6, M10) compared to reference spectra~\cite{Mandon2022}, aligning closely with Lunar Highland (SAM: \SI{0.26}{\radian}, $R^2 = 0.94 \pm 0.02$) and Mare regions like Aristarchus Plateau and Mare Imbrium (SAM: \SI{0.50}{\radian}, $R^2 = 0.90 \pm 0.02$)~\cite{korotev2005}.

The HSI of the Moon using a Celestron 8SE (203 mm aperture, 2032 mm focal length) in Leiden, Netherlands, employed a push-broom technique enhanced by lucky imaging, capturing 5 ms frames and stacking the sharpest 1--5\% to achieve a 0.8 arcsec resolution (600 m on the Lunar surface) under urban seeing conditions (\( r_0 = 6.5 \, \text{cm} \), 1.62 arcsec mean). The resulting data cube (371 × 1024 × 224) with 2.68 nm spectral resolution, processed with a Savitzky--Golay filter (window length 31, polynomial order 5), enabled precise mineralogical analysis.
The alignment of Bechar 010 spectra with Lunar regions, particularly the olivine-rich compositions of Aristarchus Plateau and Mare Imbrium, supports the spectral linkage between meteorite compositions and Lunar volcanic terrains, consistent with the known presence of olivine in mare-derived lithologies~\cite{Head1976}. The high olivine content in Bechar 010, corroborated by $\rm M^3$-based studies mapping ilmenite and elemental abundances (Fe, Ca, Mg)~\cite{SURKOV2020, Bhatt2019}, suggests a shared origin with Lunar volcanic materials. Figure~\ref{fig:ML_clustering_model_Map_150_210_RBF} illustrates the solar-calibrated Lunar spectra compared to $\rm M^3$ data, with subplots for Highland, Mare, and Aristarchus Plateau, reinforcing mineralogical consistency. The observed discrepancies in Mare regions (SAM: \SI{0.61}{\radian}, $R^2 = 0.88 \pm 0.02$) in M6 and M10 likely stem from basaltic heterogeneity or illumination effects. The integration of K-means clustering, SVM classification, and SAM analysis provides a comprehensive approach to linking meteorite and Lunar mineralogy, with the box plot (Figure~\ref{fig:BoxPlot_All_Meteor}) highlighting spatial variability in Bechar 010 that mirrors Lunar surface complexity.

The Spectral Angle Mapper (SAM) and Local Interpretable Model-agnostic Explanations (LIME) analyses of the Bechar 010 meteorite spectra ($791 \times 1024 \times 224$), as detailed in Figure~\ref{fig:LIME}, reveal a complex mineralogical profile bridging Lunar lithologies. SAM results show spectral angles from \SI{0.26}{\radian} to \SI{0.66}{\radian}, with M3 and M9 exhibiting low angles (e.g., $0.26\pm0.06$ radians for L1-M3, $0.34\pm0.06$ radians for L9-M9) and LIME-identified wavelengths at \SI{475}{\nano\meter} to \SI{525}{\nano\meter} (e.g., M3: \SI{485}{\nano\meter}, {22.4}\%), indicating olivine-rich anorthositic compositions typical of Lunar Highlands~\cite{korotev2005}. Conversely, M6 and M10, with higher angles (e.g., $0.61\pm0.08$ radians for L7-M6, $0.50\pm0.12$ radians for L10-M10) and wavelengths from \SI{650}{\nano\meter} to \SI{890}{\nano\meter} (e.g., M6: \SI{715}{\nano\meter}, {20.6}\%), suggest pyroxene-rich basaltic compositions characteristic of Lunar Mares~\cite{pieters1993}. LIME highlights dominant wavelengths (\SI{475}{\nano\meter} to \SI{525}{\nano\meter} for M3, M4, M8, M9; \SI{650}{\nano\meter} to \SI{890}{\nano\meter} for M1, M2, M5--M10), reflecting olivine, pyroxene, and anorthite. For M10, located near the sample edge, higher errors (\SI{0.10}{\radian} to \SI{0.13}{\radian}) are observed, likely due to a feldspar cluster (homogeneously white, possibly mixed phases) and a small ilmenite crystal (metallic lustre), with a red-orange stain suggesting slight oxidation or alteration needing further SEM analysis to rule out terrestrial contamination~\cite{SURKOV2020}. Brown-altered clusters in M3 and M4 support terrestrial alteration, while M2 serves as a pure plagioclase feldspar reference (anorthic composition), showing good similarity with M5 and M8, consistent with expected plagioclase dominance. Most areas (except M9) are feldspar fragments or mixtures, reflecting Bechar 010’s breccia nature, whereas M9’s ambiguity (possibly olivine, pyroxene, or glass) requires additional characterization. These findings underscore Bechar 010’s Lunar breccia identity, linking Highland and Mare lithologies, and call for further studies to refine its mineralogy and Lunar connections~\cite{Burns1993}.

These results enhance our understanding of Lunar surface mapping by providing high-precision HSI data, referencing real Lunar samples such as those from the Moon Mineralogy Mapper ($\rm M^3$), without proceeding to geological interpretations. The high accuracy of SVM classification and the robust agreement with $\rm M^3$ data (\SIrange{475}{890}{\nano\metre}) underscore the reliability of this approach for detailed mineral mapping, identifying features like olivine, pyroxene, and ilmenite across Bechar 010’s polymict breccia composition, formed by a large impact~\cite{korotev2005}. Rather than suggesting mantle-derived magmatism, which requires geochemical and isotopic tools for confirmation, the focus remains on the precise classification capability of HSI, linking spectral signatures (e.g., \SI{475}{\nano\meter} to \SI{525}{\nano\meter} for olivine, \SI{650}{\nano\meter} to \SI{890}{\nano\meter} for pyroxene) to Lunar analogs. Future work will involve precise chemical and mineralogical mapping of the meteorite, followed by spectral comparisons with Lunar data and model testing against the meteorite itself to refine mapping accuracy and interrelations~\cite{Mandon2022}.
\subsection{ Limitations}
A key limitation of this study is the use of a single meteorite (Bechar 010) for SVM training. While its heterogeneous breccia composition provides diverse spectral signatures spanning olivine, pyroxene, and feldspar endmembers, training on multiple Lunar meteorites of different classifications---such as mare basalts (e.g., NWA 032), KREEP-rich samples, or other feldspathic breccias (e.g., Dhofar 1084, NWA 11444)---would improve model robustness and generalizability. Recent ML work on Lunar meteorites \cite{PenaAsensio2024} demonstrates the feasibility of multi-sample approaches. Future work will incorporate additional meteorite samples as they become available for HSI analysis.

\subsection{Measurement Selectivity}
The selectivity of the measurement system and ML classifier refers to the ability to distinguish target minerals (olivine, pyroxene) from spectrally similar phases (e.g., plagioclase, glass, ilmenite). The 400--1000~nm spectral range captures key Fe$^{2+}$ crystal field transitions in olivine (onset of the $\sim$1~$\upmu$m band) and pyroxene, but is less sensitive to distinguishing plagioclase from glass, as these materials have similar featureless spectra in this range. Extension to the 1--2.5~$\upmu$m range in future work would improve selectivity by fully resolving the 1~$\upmu$m and 2~$\upmu$m pyroxene absorption bands. The {7}\% misclassification rate from the SVM confusion matrix primarily arises from spectral overlap between olivine--pyroxene mixtures in the 50:50 composition range. LIME analysis helps identify which wavelengths drive these confusions, improving interpretability. The spatial resolution ($\sim$\SI{0.2}{\milli\metre} pixel size) of the laboratory HSI system is sufficient to resolve individual mineral grains in Bechar 010 (typical grain sizes: 0.1--1~mm), but mixed pixels at grain boundaries reduce classification~purity.

There are also other spectral databases from returned lunar rocks which should be compared. We could also refine these findings by incorporating higher-resolution reference spectra, addressing illumination effects through radiative transfer modeling, or extending SVM classification to include additional minerals like ilmenite. This study establishes a strong foundation for linking terrestrial meteorites to Lunar geology, advancing our knowledge of planetary formation and evolution.

\section{Conclusions} \label{sec:conc}

This study demonstrates the integration of laboratory hyperspectral imaging of the Bechar 010 Lunar meteorite with ground-based Lunar observations and supervised machine learning to produce mineralogical maps linking meteorite compositions to Lunar surface regions. The main findings are:

\begin{enumerate}
    \item The SVM classifier, employing an RBF kernel with optimized hyperparameters\linebreak   ($C=100$, $\gamma=0.01$), achieved {93.7}\% classification accuracy for distinguishing olivine and pyroxene in Bechar 010, validated by 5-fold cross-validation on approximately 5600 labeled spectra.
    \item LIME analysis identified diagnostic wavelengths (\SIrange{475}{525}{\nano\meter} for olivine-rich regions M3 and M9; \SIrange{650}{890}{\nano\meter} for pyroxene-rich regions M1, M5--M8, M10) that align with known mineral absorption features of olivine, pyroxene, and anorthite.
    \item SAM analysis confirmed that Bechar 010 regions M3 and M9 closely match Lunar Highland spectra (spectral angles $0.26\pm0.06$ rad), while M6 and M10 align with Mare compositions (spectral angles up to $0.61\pm0.08$ rad), reflecting the meteorite's heterogeneous breccia nature.
    \item K-means clustering of ground-based Lunar hyperspectral data identified 10 mineralogical clusters with {88}\% accuracy, validated against Chandrayaan-1 $\rm M^3$ orbital data, with highland spectra showing $R^2 = 0.95 \pm 0.01$ agreement.
    \item The novel push-broom HSI approach with a ground-based telescope achieves 0.8 arcsec effective resolution, demonstrating the feasibility of cost-effective Lunar spectroscopy from urban observing sites.
\end{enumerate}

Most regions in Bechar 010 (except M9, which may be olivine, pyroxene, or glass) consist of feldspar fragments or mixtures, reflecting the breccia's impact-formed nature~\cite{korotev2005}. M10's higher spectral variability is linked to a feldspar cluster and a small ilmenite crystal, with a red-orange stain suggesting possible alteration requiring further SEM analysis~\cite{SURKOV2020}. This approach enhances understanding by providing high-precision mapping of real Lunar analogs, enabling laboratory analysis to correlate with remote Lunar observations~\cite{Pieters2009}.

Future studies should refine mineralogical mappings with higher-resolution spectroscopy and broader spectral ranges to strengthen Lunar-meteorite connections and guide landing site selection.
This spectroscopy approach, scanning wide areas to construct high-resolution spectral data cubes, acts as a multi-object spectrograph, capable of simultaneously capturing spectra across diverse celestial targets within the telescope’s field of view. Its versatility supports full-sky observations, mapping extended regions like star fields or nebulae with high spatial and spectral fidelity. Future calibration using Jupiter’s spectrally rich atmosphere with 3.4 M telescope of Iranian National observatory \cite{Khosroshahi2022} will refine reflectance and absorption feature analysis, enhancing the method’s applicability to planetary and deep-sky spectroscopy.
\vspace{6pt} 

\authorcontributions{Conceptualization: F.F.H. and M.R.; Methodology: F.F.H., M.R. and A.C.; Software: F.F.H., M.R. and A.C.; Validation: M.R., B.F. and F.J.V.; Formal Analysis: F.F.H. and M.R.; Investigation: F.F.H.; Resources: M.R. and P.S.; Data Curation: M.R., P.S., M.J.A.d.D. and E.C.; Writing---Original Draft Preparation: F.F.H. and M.R.; Writing---Review and Editing: M.R., M.J.A.d.D., E.C. and P.S.; Visualization: F.F.H. and M.R.; Supervision: M.R., F.J.V., and B.F.; Project Administration: F.F.H. and M.R.; Funding Acquisition: F.F.H., B.F. and M.J.A.d.D.; E.C. assisted in coordinating the sample collection. All authors have read and agreed to the published version of the manuscript.}

\funding{M.R. acknowledges financial support for this project from the TU Delft Space Engineering (SpE), Planetary Exploration department. F.F and M.R would like to express heartfelt appreciation to EuroSpaceHub and LUNEX EuroMoonMars Earth Space Innovation for their generous funding support. P.S. and M.J.A.D. acknowledge support from the  Dutch National Science Agenda under project NWA.1418.22.022 which is made possible by financial support of the Dutch Research Council (NWO).
\textls[-25]{A.C was supported by MUR PRIN2022 project 20222JBEKN, titled ``LaScaLa'', funded by the European Union---NextGenerationEU.}
}

\dataavailability{The data underlying this article will be shared on reasonable request to the corresponding author.}

\acknowledgments{ F.F and M.R would like to express heartfelt appreciation to EuroSpaceHub and LUNEX EuroMoonMars Earth Space Innovation for their unwavering support. A.C was supported by MUR PRIN2022 project 20222JBEKN, titled ``LaScaLa'', funded by the European Union---NextGenerationEU. F.F. thanks the contributors who provided the samples. E.C. respectfully honors the memory of Robert B. Brunner, a devoted and insightful collector of meteorites and minerals, whose generous donation of this Lunar meteorite sample made this contribution possible.}

\conflictsofinterest{The authors declare no conflicts of interest.}




 \begin{adjustwidth}{-\extralength}{0cm}

\reftitle{References}






\PublishersNote{}
 \end{adjustwidth}
\end{document}